\documentclass[twocolumn,times,tighten]{aastex62}
\usepackage{graphicx}		% for including figures
\usepackage{latexsym}		% additional math symbols
\usepackage{bm}			% bold math package
\usepackage[english]{babel}
\usepackage{color}		% for colored text and boxes; remove for subm.
\newcommand{\lsim}{\mbox{\rlap{\hbox{\lower2pt\hbox{\ensuremath{\sim}}}}\raise2pt\hbox{\ensuremath{<}}}}%
\newcommand{\gsim}{\mbox{\rlap{\hbox{\lower2pt\hbox{\ensuremath{\sim}}}}\raise2pt\hbox{\ensuremath{>}}}}%
\renewcommand{\gtrsim} {\ensuremath{\gsim}}
\renewcommand{\lesssim}{\ensuremath{\lsim}}
\renewcommand{\farcs}  {\mbox{\ensuremath{.\mkern-5mu^{\prime\prime}}}}%
\newcommand{\eg}	{e.g., }%
\newcommand{\ie}	{i.e., }%
\newcommand{\etal}	{\hbox{et~al.}}%

\newcommand{\HST}	{\emph{HST}}%
\newcommand{\JWST}	{\emph{JWST}}%
\newcommand{\Spitzer}	{\emph{Spitzer}}%
\newcommand{\WISE}	{\emph{WISE}}%
\newcommand{\Euclid}	{\emph{Euclid}}%
\newcommand{\WFIRST}	{\emph{WFIRST}}%
\newcommand{\Chandra}	{\emph{Chandra}}%

\newcommand{\EBV}	{\textsl{E(B$-$V)}}%
\newcommand{\mAB}	{\ensuremath{m_{\rm AB}}}%
\newcommand{\muJy}	{\ensuremath{\mu{\rm Jy}}}%
\newcommand{\tsim}	{\ensuremath{\sim}}%
\newcommand{\tsimeq}	{\ensuremath{\simeq}}%
\newcommand{\ttimes}	{\ensuremath{\times}}%
\newcommand{\tpm}	{\ensuremath{\pm}}%
\newcommand{\tdeg}	{\ensuremath{^{\circ}}}%
\newcommand{\tmin}	{\ensuremath{'}}%

\newcommand{\code}[1]	{\textsf{\small #1}}%		# for in main text
\newcommand{\scode}[1]	{\textsf{\scriptsize #1}}%	# for in captions
\newcommand{\swkey}[1]	{\textsf{\footnotesize #1}}%	# keywords in software

\newlength{\txw}
\newlength{\txh}

\begin{document}
%%%%%%%%%%%%%%%%%%%%%%%%%%%%%%%%%%%%%%%%%%%%%%%%%%%%%%%%%%%%%%%%%%%%%%%%%%%%

% Paper I: Jansen & Windhorst 2018a:
\title{\Large\bfseries The \emph{James Webb Space Telescope} North Ecliptic
	Pole Time-Domain Field -- I:\\Field Selection of a \JWST\ Community
	Field for Time-Domain Studies}

% Paper II in this series (in prep.):
% Jansen, Ashcraft, Willmer, Windhorst, Cohen, \etal\ 2019a:
%\title{The \JWST\ North Ecliptic Pole Time-Domain Field -- II: \textsl{Ugriz}
%     Source Photometry to \mAB\lesssim26.X\,mag with the Large Binocular
%     Cameras}

% Paper III in this series (in prep.):
% Jansen, Grogin, Windhorst, Koekemoer, Cohen, \etal\ 2019b:
%\title{The \JWST\ North Ecliptic Pole Time-Domain Field -- III: UV--visible
%     Source Photometry and Characterization with the \emph{Hubble Space       
%     Telescope} Wide Field Camera 3 and Advanced Camera for Surveys}

%%\author[orcidID]{Author Name}
\author[0000-0003-1268-5230]{Rolf A. Jansen}
\author[0000-0001-8156-6281]{Rogier A. Windhorst}

\affiliation{School of Earth \& Space Exploration, Arizona State University,
	Tempe, AZ\,85287-1404, USA}

\correspondingauthor{Rolf A. Jansen}
\email{Rolf.Jansen@asu.edu}

\journalinfo{Publications of the Astronomical Society of the Pacific --- 
	draft version \today}

\vspace*{-12pt}
\received{2018 July 13}
%\revised{\today}
\accepted{2018 September 24}
%\published{\today}
%\submitjournal{PASP}

%\watermark{Embargoed draft}
%\setwatermarkfontsize{1in}

\shortauthors{Jansen \& Windhorst}
\shorttitle{The \JWST\ NEP Time-Domain Field -- I}

%% Abstract must be a single paragraph of not more than 250 words
\begin{abstract}
We describe the selection of the \emph{James Webb Space Telescope} (\JWST)
North Ecliptic Pole (NEP) Time-Domain Field (TDF), a \gsim14\tmin\ diameter
field located within \JWST's northern Continuous Viewing Zone (CVZ) and
centered at (RA, Dec)$_{\rm J2000}$ = (17:22:47.896, +65:49:21.54).  We
demonstrate that this is the \emph{\bfseries only} region in the sky where
\JWST\ can observe a clean (\ie free of bright foreground stars and with
low Galactic foreground extinction) extragalactic deep survey field of this
size \emph{\bfseries at arbitrary cadence} or at \emph{\bfseries arbitrary
orientation}, and without a penalty in terms of a raised Zodiacal
background.  This will crucially enable a wide range of new and exciting
time-domain science, including high redshift transient searches and
monitoring (\eg SNe), variability studies from Active Galactic Nuclei (AGN)
to brown dwarf atmospheres, as well as proper motions of possibly extreme
scattered Kuiper Belt and Inner Oort Cloud Objects, and of nearby Galactic
brown dwarfs, low-mass stars, and ultracool white dwarfs.  A
\JWST/NIRCam+NIRISS GTO program will provide an initial
0.8--5.0\micron\ spectrophotometric characterization to
\mAB\,\tsim\,28.8\tpm0.3\,mag of four orthogonal ``spokes'' within this
field.  The multi-wavelength (radio through X-ray) context of the field is
in hand (ground-based near-UV--visible--near-IR), in progress (VLA 3\,GHz,
VLBA 5\,GHz, \HST\ UV--visible, \Chandra\ X-ray, IRAM\,30m 1.3 and 2\,mm),
or scheduled (JCMT 850\micron).  We welcome and encourage ground- and
space-based follow-up of the initial GTO observations and ancillary data,
to realize its potential as an ideal \JWST\ time-domain
\emph{\bfseries community field}.
\end{abstract}

%*% Select up to six keywords; they must be listed in alphabetical order.
%*% (there exists no time-domain related keyword in the current AAS list)
% remove "--- time-domain science" when submitting the manuscript

\section{Introduction}

In an age when surveys with the Panoramic Survey Telescope and Rapid
Response System \citep[Pan-STARRS;][]{PanSTARRS}, Large Synoptic Survey
Telescope \citep[LSST;][]{LSST}, \Euclid\ \citep{Euclid}, and Wide Field
Infrared Survey Telescope \citep[\WFIRST;][]{WFIRST} are or will soon allow
time domain studies of relatively faint objects within our Solar System,
Galactic neighorhood and beyond, as well as at cosmological distances and
associated large look-back times, one may ask to what extent the \emph{James
Webb Space Telescope} \citep[\JWST;][]{JWST,JWST2} can serve as a
time-domain survey facility.  Whereas LSST's limit for variability studies
over a large portion of the sky in the near-UV--near-IR reaches to {\mAB}
{\lsim} 24 mag (10\,$\sigma$) per 2\ttimes15\,s visit on time-scales of
\tsim15\,min--1\,hour \citep{LSST,LSSTaas14,LSSTobs17}, \JWST\,/\,NIRCam
\citep[e.g.,][]{NIRCam1,NIRCam2} could potentially reach {\mAB} {\tsim}
26.8--28.3 mag (10\,$\sigma$) per epoch in the near--mid-IR on similar
time-scales in a suitably dark survey field\footnote{As reported by the
\JWST\ ETC \citep{JWSTetc} v1.2 available at
\url{https://jwst.etc.stsci.edu/}.}.
  %
%*% Note: LSST depth is usually quoted at 5 sigma: ~24 [Tyson 2002] or ~24.5
%*%   [Ivezic 2014] for back-to-back 15s exposures in r.  The latter would
%*%   reach no deeper than 23.75 mag at 10 sigma, so "<~24 mag" is generous.
%*%
%*% Note: JWST ETCv1.2 https://jwst.etc.stsci.edu/
%*%   (see also: Pontoppidan et al. 2016, SPIE 9910, 16)
%*%   NIRCam SW F200W point source imaging and a read-out
%*%   pattern of MEDIUM8,5grp/int,7ints --> t_exp=3682.7 s:
%*%     point source, m_AB=29.0 (band-averaged) --> S/N =  5.10 +/- 0.08
%*%     point source, m_AB=28.2 (band-averaged) --> S/N = 10.45
%*%   pattern of SHALLOW4,4grp/int,3ints --> t_exp=644.2 s:
%*%     point source, m_AB=27.0 (band-averaged) --> S/N =  8.45
%*%     point source, m_AB=26.8 (band-averaged) --> S/N = 10.07
%*%   NIRCam SW F200W, z=0 point source with flat spectrum in f_nu and a
%*%   read-out pattern of DEEP8,3grp/int,1int --> t_exp=526.10 s:
%*%     unrecoverable (1st group) saturation in >=1 pixel at m_AB=19.80mag
%*%   read-out pattern of MEDIUM8,3grp/int,1int --> t_exp=311.37 s:
%*%     unrecoverable (1st group) saturation in >=1 pixel at m_AB=19.25mag
%*%   read-out pattern of SHALLOW4,3grp/int,1int --> t_exp=161.05 s:
%*%     unrecoverable (1st group) saturation in >=1 pixel at m_AB=18.45mag
  %
This would enable a wide range of new and exciting time-domain science in an
unexplored magnitude regime, including high redshift transient searches and
monitoring (\eg Type Ia SNe to $z$\,\tsim\,5 and Core Collapse SNe to
$z$\,\tsim\,1.5 \citep{Grauretal14,Rodneyetal14,Rodneyetal15b,Strolgeretal15b,
Mesingeretal06} and Pair Instability SNe in the Epoch of Reionization
\citep{PanKasenLoeb12,Gal-Yam12,Nicholletal13,Whalenetal13,Whalenetal14};
variability studies from (weak) Active Galactic Nuclei \citep[AGN;
e.g.,][]{QSOvarSDSS82} to brown dwarf atmospheres
\citep{Artigauetal09,Buenzlietal14,Radiganetal14,Rajanetal15};
as well as perhaps proper motions of extreme scattered Kuiper Belt, inner
Oort Cloud Objects and comets on their way in toward the inner Solar System,
and of nearby Galactic brown dwarfs and low-mass stars
\citep{Ryanetal11,RyanReid16}, and ultracool white dwarfs
\citep{Harrisetal06,Catalanetal13}. 
  %
%%%%%%%%%%%%%%%%%%%%%%%%%%%%%%%%%  FIGURE 1  %%%%%%%%%%%%%%%%%%%%%%%%%%%%%%%%
%*% Note: 'pdflatex' trips on file names with multiple dots; workaround is to
%*%       place everything other than the extension between curly brackets.
\noindent\begin{figure*}[ht]
\centerline{
  \includegraphics[width=0.489\txw]{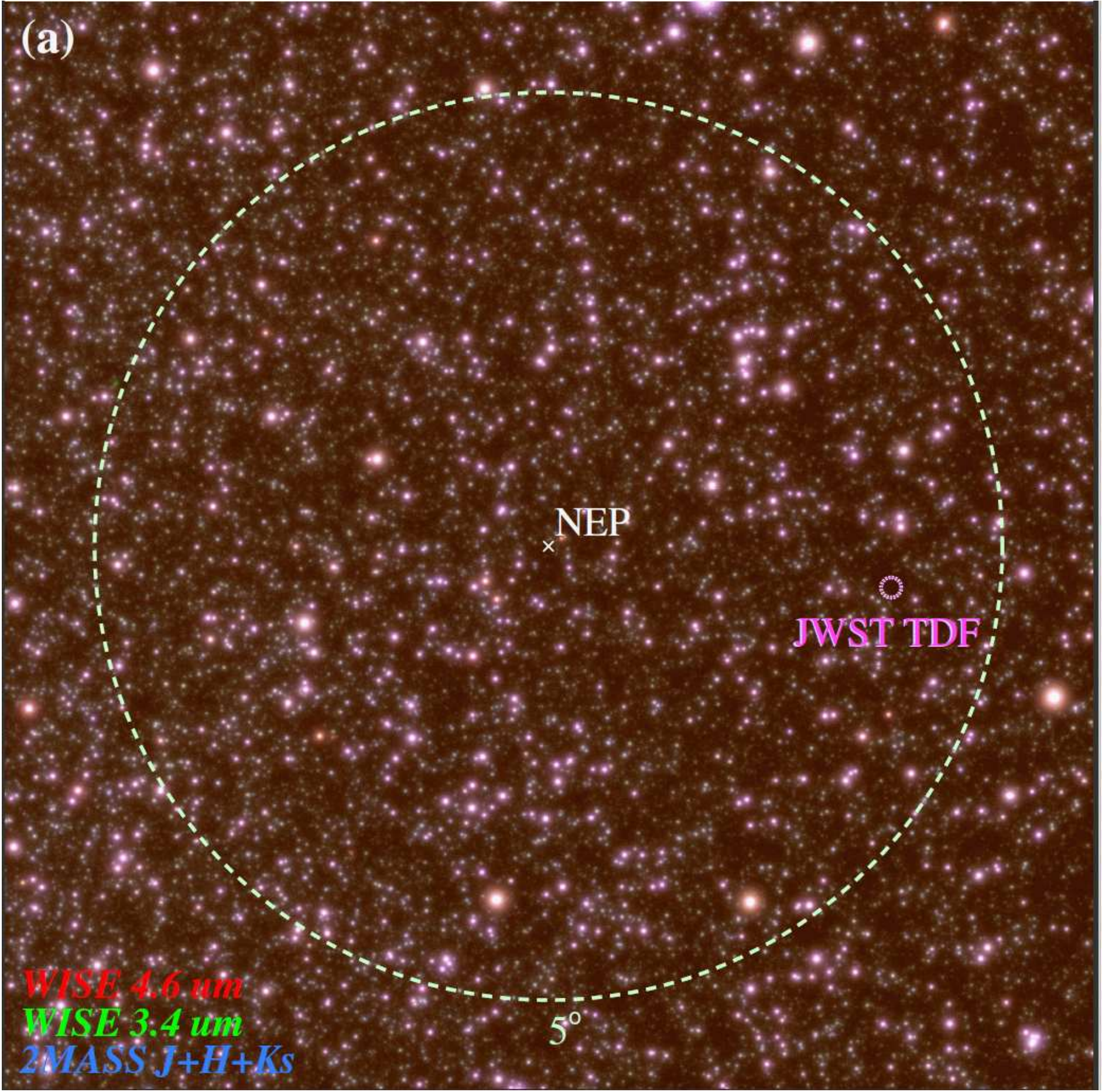}
  \hspace*{1mm}
  \includegraphics[width=0.489\txw]{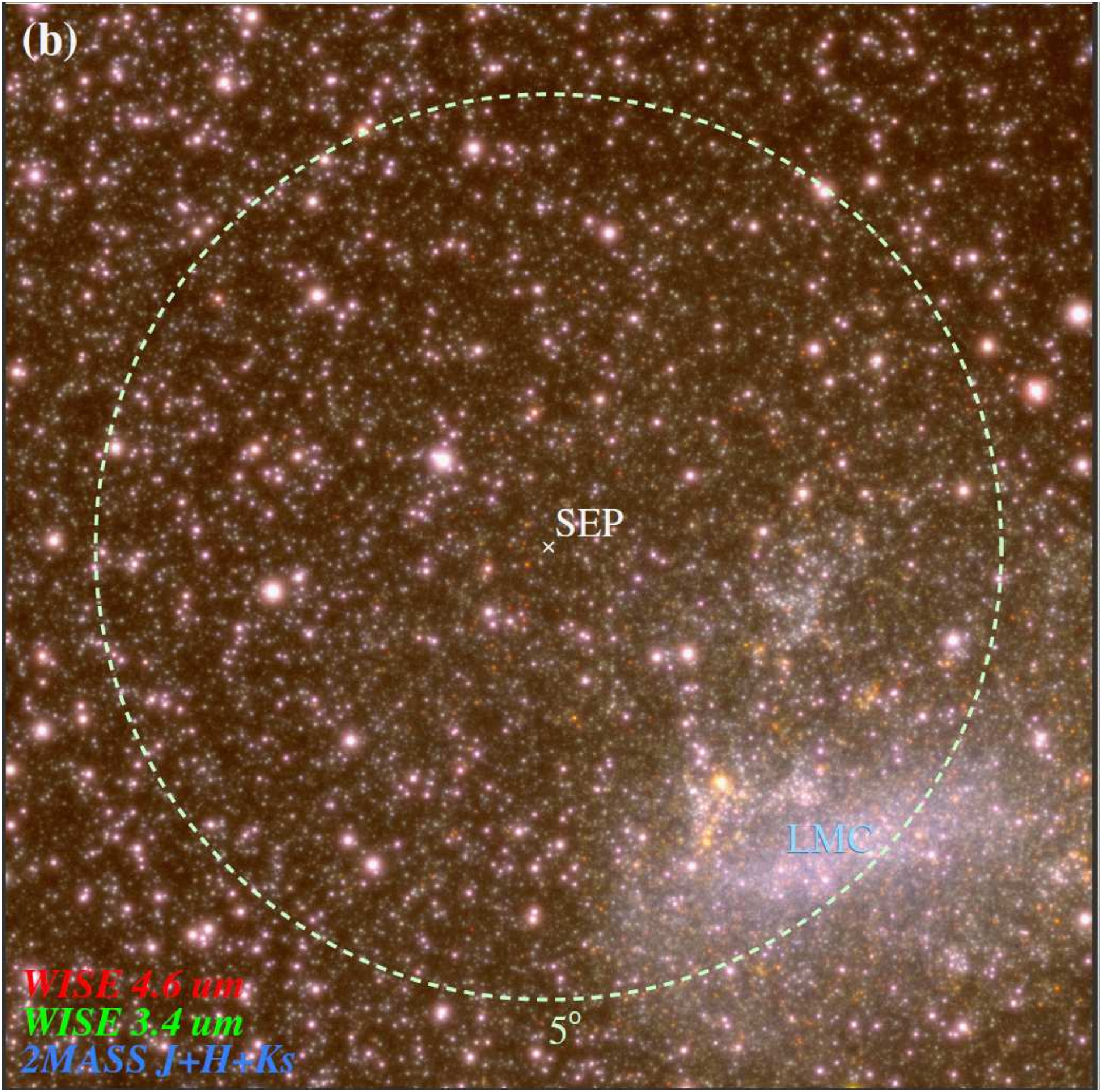}
}
\caption{Near-infrared color composites of \JWST's [\emph{a}] northern, and
[\emph{b}] southern continuous viewing zone (CVZ; indicated by the pale green
dashed circle with a radius of 5\tdeg).  In each 12\tdeg\ttimes12\tdeg\ 
image, 2MASS \emph{J+H+K$_{\rm s}$}, \WISE\ 3.4\micron, and \WISE\ 
4.6\micron\ images are shown in blue, green, and red hues.  For display
purposes only, to better perceive source over-densities and relative
brightnesses over such a wide area of sky, the constituent images were
smoothed to an effective core resolution of \tsim1.3\tmin, with a halo of
\tsim5.7\tmin\ (FWHM).  The stretch of the images contributing to each of
the two color composites is identical. \JWST's southern CVZ is dominated by
the Large Magellanic Cloud (LMC) and a denser distribution of Galactic
stars, rendering it less suitable for deep extragalactic surveys with
\JWST/NIRCam, while the northern CVZ has portions that appear relatively
empty of sources that are bright at these wavelengths. The very best region
selected in the northern CVZ (see \S\,3) is indicated by a small, dotted,
magenta circle with a radius of 7{\tmin} and labeled ``JWST TDF''.
\label{fig:jwstCVZs_nir}}
\end{figure*}
%%%%%%%%%%%%%%%%%%%%%%%%%%%%%%%%%%%%%%%%%%%%%%%%%%%%%%%%%%%%%%%%%%%%%%%%%%%%%
  %
%%%%%%%%%%%%%%%%%%%%%%%%%%%%%%%%%  FIGURE 2  %%%%%%%%%%%%%%%%%%%%%%%%%%%%%%%%
\noindent\begin{figure*}[ht]
\centerline{
  \includegraphics[width=0.489\txw]{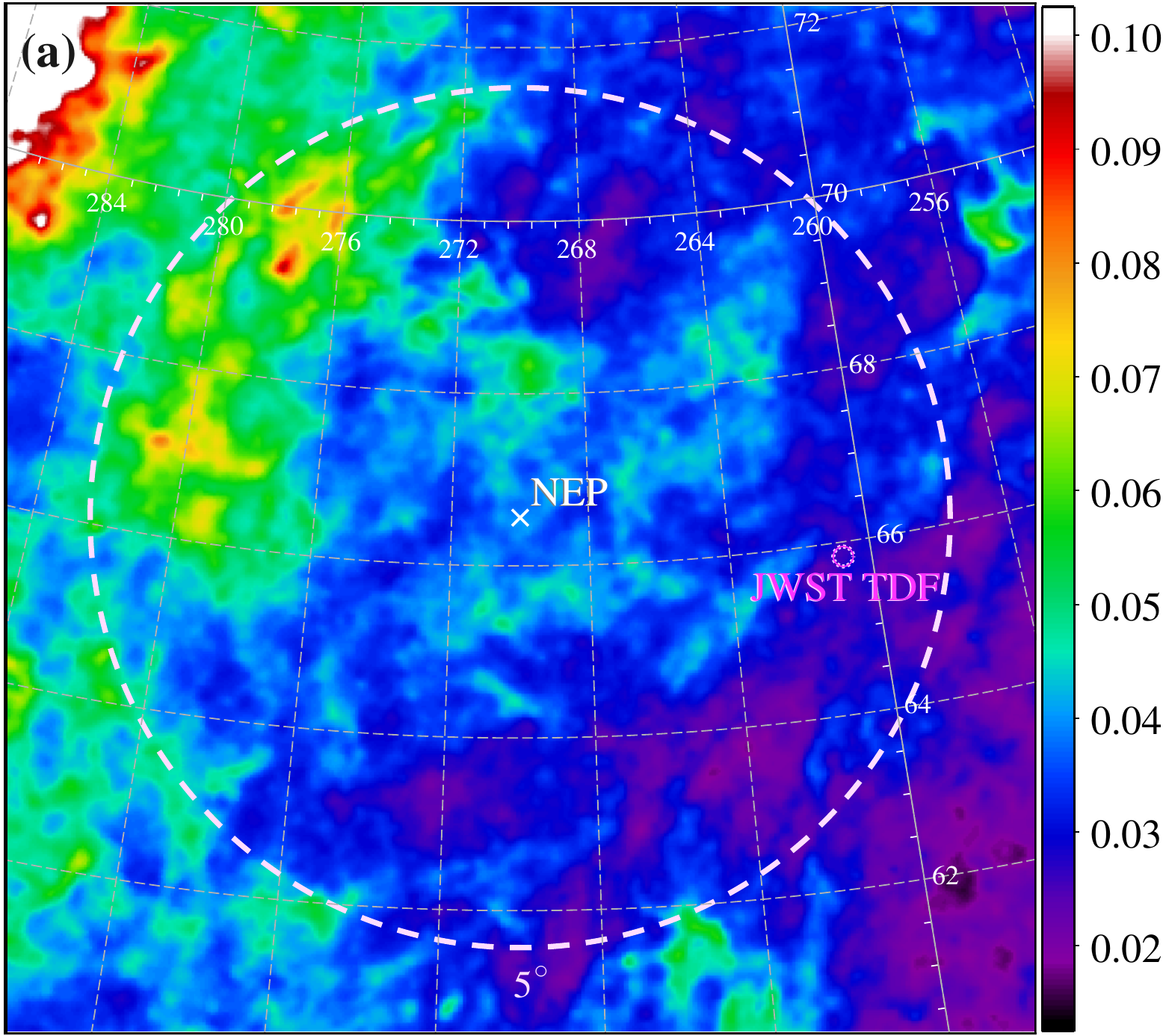}
  \hspace*{1mm}
  \includegraphics[width=0.489\txw]{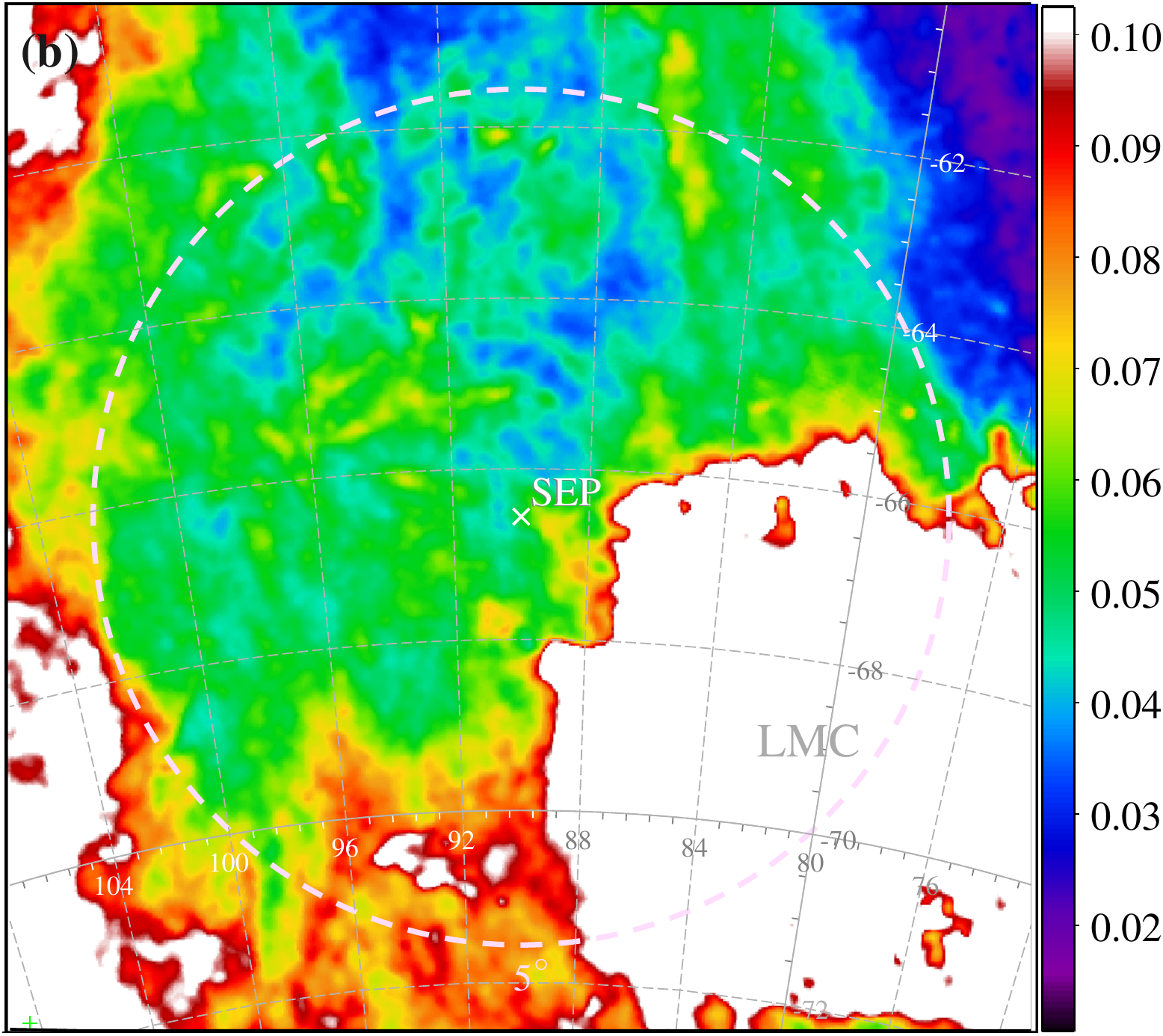}
}
\caption{\noindent\small
Map of \EBV\ values in the 12\tdeg\ttimes12\tdeg\ area around (\emph{a}) the
North Ecliptic Pole, and (\emph{b}) the South Ecliptic Pole, adopting the
\citet{SF11} recalibration of the \citet{SFD98} Galactic dust reddening map.
As in Fig.~\ref{fig:jwstCVZs_nir}, the dashed circles with a radius of
5\tdeg\ indicate \JWST's Northern and Southern Continuous Viewing Zones. The
location of the \JWST\ Time-Domain Field (see \S\,3) is indicated by a
magenta circle with a radius of 7\tmin\ in panel (\emph{a}), while the LMC
(labeled) and other Galactic structures occupy a large portion of the
Southern CVZ in panel (\emph{b}).  Both panels are shown at identical stretch
(0.01 $\lesssim$ \EBV\ $\lesssim$ 0.10 mag).
\label{fig:jwstCVZs_ebv}}
\end{figure*}
%%%%%%%%%%%%%%%%%%%%%%%%%%%%%%%%%%%%%%%%%%%%%%%%%%%%%%%%%%%%%%%%%%%%%%%%%%%%%

Shortly after insertion of the \JWST\ in an Earth-Sun L2 halo orbit in
2021, it will start its program of Early Release Science (ERS)
observations, Guaranteed Time Observations (GTO) by the instrument teams
and Interdisciplinary Scientists (IDS), and General Observer (GO)
observations.  For most locations on the sky, Sun avoidance, power
generation, and shielding requirements of the cryogenic telescope combine
to restrict object visibility to two time intervals per year (with a
duration that depends on ecliptic latitude), except within \tsim5\tdeg\ of
the Ecliptic Poles, where \JWST\ will have two continuous viewing zones
(CVZs) in which targets are observable year-round.  Of these two small
regions in the sky, the North Ecliptic Pole (NEP) CVZ constitutes
relatively more empty sky; the Large Magellanic Cloud (LMC) and
(associated) large-scale Galactic structures occupy much of the area within
the southern CVZ, as can be seen by comparing the near-IR color composites
of Fig.~\ref{fig:jwstCVZs_nir}\emph{a,b} and the maps of Galactic
foreground reddening of Fig.~\ref{fig:jwstCVZs_ebv}\emph{a,b}.

The orientation of \JWST's instrument apertures on the sky will be dictated
by the date of observation, with limited schedulability of off-nominal
angles ($\Delta$PA$<$15\tdeg) due to sunshield and solar panel constraints.
This is in stark contrast to observations with the \emph{Hubble Space
Telescope} (\HST), where observers can find viable solutions for a very wide
range of instrument aperture orientations.  In particular, it has become
common practice to revisit targets with \HST\ rotated over 180\tdeg\ to
secure observations with large areas of overlap between a primary and a
coordinated parallel instrument at the two orientations \citep[e.g., 
CANDELS;][]{CANDELS1,CANDELS2}.

Moreover, off-nominal orientations with \JWST\ will also carry a penalty of
an increased Zodiacal background, which is up to \tsim1.2--1.4\,mag brighter
in the ecliptic plane at 1.25--4.9\,\micron\ than at the ecliptic poles
\citep{COBEDIRBE2} ---\,and hence results in a penalty of reduced
sensitivity per unit observing time.  Here also, the \JWST\ CVZs are the
exception, where any desired orientation can be scheduled at some time
during the year, and the Zodiacal background is always at a minimum. 
Residual variation in the background signal in the CVZs is mostly due to
orientation-dependent straylight from the Milky Way, since its combined
starlight can reflect off the rear surface of \JWST's sunshield and enter
the science instrument apertures along unintended paths
\citep{JWSTstray1,JWSTstray2}.  In general, the level of straylight may
reach up to \tsim40\% of the Zodiacal background.

%*% Note1: JWST ETCv1.2: variation in overall background throughout the
%*%   year at the location of the JWST NEP-TDF is +/- 0.08 in terms of S/N
%*%   achieved in a given exposure time in NIRCam SW F200W.  This must be
%*%   mostly due to modeled stray MW light and not due to Zodi.
%*% Note2:
%*%   NEP is at (l_e,b_e) = (0,+90)  --> (l^II,b^II) ~ ( 96.38,+29.81)
%*%   SEP is at (l_e,b_e) = (0,-90)  --> (l^II,b^II) ~ (276.38,-29.81)
%*%   Galactic ctr is at (l^II,b^II)=(0,0) --> (l_e,b_e) ~ (266.84,-5.54)
%*%   Galactic anti-ctr is at      (180,0) --> (l_e,b_e) ~ ( 86.84,+5.54)
%*%   JWST NEP TDF is at (l^II,b^II) = (95.8394658, +33.6058400)
%*%                    --> (l_e,b_e) = (186.8417287,+86.1775634)

Given that \JWST's CVZs are located at intermediate Galactic latitudes
($b^{\rm II}$\,\tpm\,\tsim\,30\tdeg), a field that would be optimal for a
deep extragalactic time-domain survey with \JWST\ would also be suitable
for deep Galactic time-domain science, sampling stellar and sub-stellar
populations in the nearby thin and thick disk, and stars in the more
distant halo of our Galaxy.  Very faint brown dwarfs and late-type low-mass
stars will be detectable with \JWST/NIRCam through\linebreak\newpage

\noindent both their near-infrared
colors and proper motions, where the exquisite resolution of \JWST\ ensures
robust star--galaxy separation.

Little is known of objects in the very outer realms of our Solar System at
very high ecliptic latitudes near \tpm90\tdeg, should they exist. The
deepest and widest high ecliptic latitude survey of trans-Neptunian objects
to date \citep[701\,deg$^2$ to $r_{\rm AB}$ \lsim\ 22.4--24.8 
mag;][]{CFEPS-HL} indicates that $i$\,\tsim\,90\tdeg\ objects exist but are
rare. Their size distribution may be flatter, but their albedo higher and
color bluer, than that typical of objects in the dynamically cold population
near the ecliptic plane \citep[e.g.,][and references therein]{PDSSS,
OSSOS-II, CFEPS-HL}.  The Oort Cloud \citep{Oort1950} is assumed to form a
roughly spherical distribution of objects that orbit the Sun at distances
beyond 2000\,AU out to at least \tsim50,000\,AU.  These left-overs from the
birth of our Solar System and objects captured from other star systems
during close passages over the past 4.6\,Gyr, if perturbed and on an inward
trajectory, may be detectable with \JWST\ at 27\,\lsim\,\mAB\,\lsim\,29\,mag
via their expected large parallaxes.  A large, relatively reflective object
like (90377)\,Sedna ($D_{\oplus}$\,=\,88.1\,AU, $d$\,\tsim\,1000\,km,
$A$\,\tsim\,0.32, $m_{V,{\rm Vega}}$\,$\simeq$\,21\,mag, and
$V\!-\!K$\,$\simeq$\,2.1\,mag; \citealt{Sedna}) has $m_{K,{\rm
AB}}$\,$\simeq$\,21.7\,mag and would be directly detectable to a distance of
\tsim500\,AU for a limiting magnitude of 29.  Assuming a darker, KBO-like
albedo of $A$\,\tsim\,0.04 \citep[\eg][]{LuuJewitt98}, a comet nucleus as
small as 10\,km in diameter on its way in from the Oort Cloud would be
detectable to a distance of 28--29\,AU, comparable with the distance to
Neptune.  Outer Solar System objects within \JWST's CVZs would display an
annual parallax, describing a circle with a radius $r = (180/\pi)(1/R)$\tdeg,
with $R$ their distance from the Sun in AU, and move \tsim148\,$R^{-1}$ 
$''$/hour due to their parallax
  %
%*% 2*pi * (180/pi) * (1/D) / (365.25*24) [deg/hr] ~ 147.84*(1/D) ["/hr]
  %
(their apparent motion as a result of their orbital velocity is more than
two orders of magnitude smaller). 
This and the low zodiacal foreground near the ecliptic poles would make the
CVZs ideally suited to search for such objects, allowing detections of
\gsim10\,km comets at the distance of Neptune even within a single
\tsim1--2\,hr visit.

In this first paper on the \JWST\ North Ecliptic Pole (NEP) Time-Domain
Field (TDF), we describe the selection of this new, very best target field
for a deep extragalactic time-domain survey with \JWST.  For the reasons
stated, we restrict our analysis to \JWST's CVZs, where such a survey could
include the time domain from time scales of \tsim10\,min to
10--14\,years, the anticipated maximum lifetime of \JWST\ 
\citep[{e.g.,}][and references therein]{Windhorstetal18}.
We further discuss considerations for a practical implementation and
development of this field as both a \JWST\ GTO program and as a GO
community field.  And we briefly list specific considerations for the
ancillary ground- and space-based observations across the electromagnetic
spectrum that have been secured or awarded to date, and that will each be
described in detail in future papers.

\section{Selection of the best field for a deep extragalactic \JWST\ survey}

\subsection{Bright object concerns}

The large \tsim6.5\,m aperture of \JWST\ and the large twin
2\farcm2\ttimes2\farcm2 fields of view of NIRCam, coupled with persistence
in its sensitive near- and mid-IR detectors, pose significant constraints
on the presence and brightness of stars and other objects that can be
tolerated within the footprint of a deep \JWST\ survey.  The persistence
acts as localized reductions in sensitivity and increases in image noise,
and forms a record of observations that may have occured hours earlier (and
hence is a source of sample contamination). For \JWST/NIRCam detector
read-out patterns \swkey{SHALLOW4}, \swkey{MEDIUM8}, and \swkey{DEEP8},
appropriate for medium-deep and deep extragalactic surveys, unrecoverable
saturation (\ie when full-well charge capacity is reached already before or
during the first group of reads in an integration) in a F200W observation
of a flat-spectrum point source can set in for sources as faint as
\mAB\,\tsimeq\,18.45, 19.25, and 19.80\,mag, respectively\footnote{As
reported by the \JWST\ ETC \citep{JWSTetc} v1.2 available at
\url{https://jwst.etc.stsci.edu/}\,.}.
At mild saturation, the effects of persistence in subsequent exposures are
expected to be fully modelable and correctable \citep{Leisenring16}.  This
will no longer be the case for objects with \mAB\,\lsim\,15.5\,mag, which
are bright enough to deeply saturate the NIRCam detectors.
The ideal field for a deep \JWST\ survey must, therefore, be devoid of any
bright (\mAB\,$<$\,15.5\,mag) stars that would cause such deep saturation.
  %
%*% Note: JWST ETCv1.2 https://jwst.etc.stsci.edu/ gives onset of saturation
%*%   for NIRCam SW F200W, z=0 point source with flat spectrum in f_nu and a
%*%   read-out pattern of DEEP8,3grp/int,1int --> t_exp=526.10 s:
%*%     unrecoverable (1st group) saturation in >=1 pixel at m_AB~19.80mag
%*%   read-out pattern of MEDIUM8,3grp/int,1int --> t_exp=311.37 s:
%*%     unrecoverable (1st group) saturation in >=1 pixel at m_AB~19.25mag
%*%   read-out pattern of SHALLOW4,3grp/int,1int --> t_exp=161.05 s:
%*%     unrecoverable (1st group) saturation in >=1 pixel at m_AB~18.45mag
%*% These numbers were independently verified by Jarron Leisenring with his
%*% own NIRCam ETC as being indeed correct (private comm.; via Christopher
%*% Willmer dd. Feb 14 2018).

\subsection{Survey Field Size and Coordinated Parallels}

If we aim to employ \HST's highly efficient survey strategy of revisiting a
pair of primary and coordinated parallel observations with the respective
instrument footprints swapped by rotating the observatory over 180\tdeg{}
---\,possible without penalty (or at all) for \JWST\ only within its two
CVZs\,---, then we also need to take into account the projected distances
between these instruments within \JWST's focal plane\footnote{See
\emph{JWST Field of View}, JWST User Documentation [updated 2018 July 1],
STScI (Baltimore, MD);
\url{https://jwst-docs.stsci.edu/display/JTI/JWST+Field+of+View}\,.}.
For NIRCam observations with parallel NIRISS observations, the
corner-to-corner angular span of these cameras as projected on the sky and,
hence, the absolute \emph{minimum} diameter of a clean survey field, is
\tsim10\farcm2.  Similarly, for NIRSpec or MIRI observations with NIRCam
parallels it would be \tsim11\farcm6 or \tsim11\farcm4, respectively.
  %
%*% Note: https://jwst-docs.stsci.edu/display/JTI/JWST+Field+of+View
%*%   Table 1 on that JDOx page (viewed dd. Feb 14 & Sep 18 2018) lists:
%*%     NIRCam  NRCALL_FULL   (V2_1,V3_1)=( 153.16",-559.27")
%*%     NIRCam  NRCALL_FULL   (V2_2,V3_2)=(-153.74",-557.14")
%*%     NIRCam  NRCALL_FULL   (V2_4,V3_4)=( 151.48",-427.95")
%*%     NIRISS  NIS_CEN       (V2_2,V3_2)=(-356.38",-765.25")
%*%     MIRI    MIRIM_ILLUM   (V2_3,V3_3)=(-485.67",-315.04")
%*%     NIRSpec NRS_FULL_MSA1 (V2_3,V3_3)=( 535.09",-438.70")
%*%   corner-to-corner NIRCAM(V2_4,V3_4)--NIRISS(V2_2,V3_2):
%*%       sqrt((151.48+356.38)**2+(-427.95+765.25)**2) = 609.67" = 10.16'
%*%   corner-to-corner NIRCAM(V2_1,V3_1)--MIRIM(V2_3,V3_3):
%*%       sqrt((153.16+485.67)**2+(-559.27+315.04)**2) = 683.92" = 11.40'
%*%   corner-to-corner NIRCAM(V2_2,V3_2)--NIRSPEC(V2_3,V3_3):
%*%       sqrt((-153.74-535.09)**2+(-557.14+438.70)**2)= 698.94" = 11.65'
%*%   This updates my earlier (anno 2016) rougher values of 10' and ~11'.
%*%   Note that small changes in the exact positions of these vertices are
%*%   expected after spacecraft I&T and, especially, after launch.

During experimentation with an areal survey layout that
combined primary NIRCam imaging with parallel NIRISS wide-field slitless
spectroscopy, we found that large dithers would be needed to span the gap
between NIRCam modules A and B, and/or to advance the survey pointing by
the width or height of a NIRISS footprint on the sky.  This is discussed in
more detail in \S\,4.  A practical survey field size would therefore need
to be significantly larger than \tsim10\farcm2.  A field with a diameter of
\tsim14\tmin\ would accommodate the imaging instruments (NIRCam, NIRISS,
and MIRI) in \JWST's focal plane, or NIRCam in combination with NIRSpec, at
any orientation, and allow contiguous survey coverage with sufficient
freedom for various dithering strategies.  Our aim is therefore to find
regions of at least this size within the \JWST\ CVZs that are devoid of
near-IR-bright sources.
  %
%%%%%%%%%%%%%%%%%%%%%%%%%%%%%%%%%  FIGURE 3  %%%%%%%%%%%%%%%%%%%%%%%%%%%%%%%%
\noindent\begin{figure*}[ht]
\centerline{
  \includegraphics[width=0.49\txw]{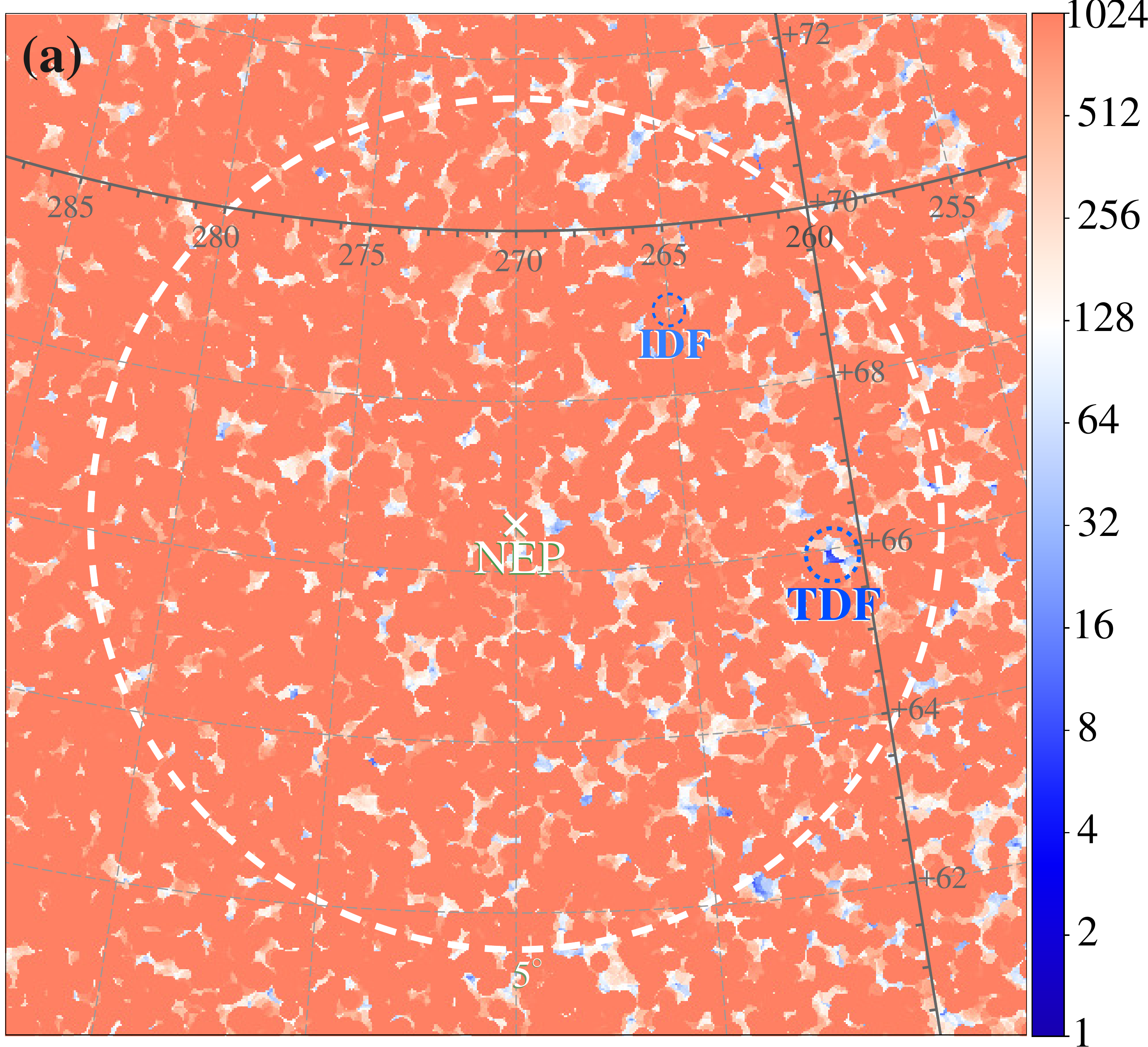}
  \hspace*{1mm}
  \includegraphics[width=0.49\txw]{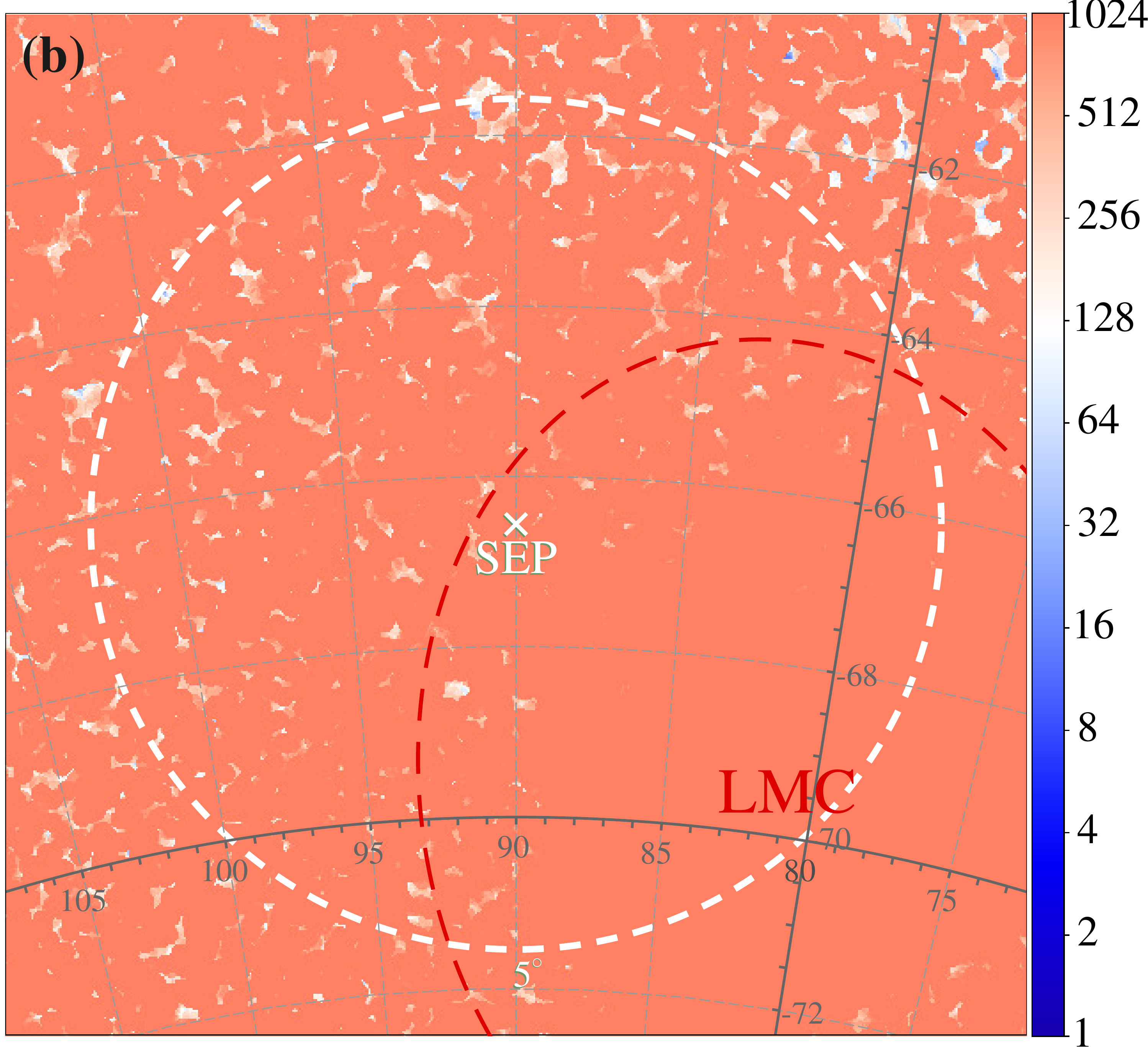}
}
\caption{\noindent\small
Map of \tsim\,4\micron\ source penalties cumulated within round, 14\tmin\
diameter viewports stepped with 20$''$ steps through the
12\tdeg\ttimes12\tdeg\ area around (\emph{a}) the North Ecliptic Pole, and
(\emph{b}) the South Ecliptic Pole. As in Fig.~\ref{fig:jwstCVZs_nir}, the
dashed circles with a radius of 5\tdeg\ indicate \JWST's Northern and Southern
Continuous Viewing Zones. The location of the \JWST\ Time-Domain Field (TDF)
and of the IRAC Dark Field (IDF; see \S\,3.3) are indicated with labeled,
dotted circles in panel (\emph{a}), while the LMC (labeled) and other Galactic
structures occupy a large portion of the Southern CVZ.  Penalties are
normalized such that 1 corresponds to the equivalent of a single
\mAB\,=\,15.5\,mag source within a viewport.  Both panels are shown at
identical stretch, focusing on penalties in the 1--1024 range (the best
14\tmin\ diameter viewports have cumulated penalties less than 10). There are
\emph{very} few regions of this size that are devoid of sources brighter than
\mAB\,=\,15.5\,mag, appearing in dark blue hues in these maps.
\label{fig:jwstCVZs_srcpen}}
\end{figure*}
%%%%%%%%%%%%%%%%%%%%%%%%%%%%%%%%%%%%%%%%%%%%%%%%%%%%%%%%%%%%%%%%%%%%%%%%%%%%%

\subsection{Moving Viewport Analysis of {\itshape WISE} 3.4+4.6\micron\ 
	Sources}

We started with an exploratory analysis of just \JWST's northern CVZ, by
shifting a 10\tmin\ttimes10\tmin\ viewport in steps of 1\tmin\ in RA and
Dec through a \tsim5.5\tdeg\ radius portion of the \WISE\ Source
Catalog\footnote{Retrieved from the NASA/IPAC Infrared Science Archive
(IRSA; \url{http://irsa.ipac.caltech.edu/}) using a custom C-shell
script, \scode{qwisesc}, to request \WISE\ and 2MASS 
\textsl{JHK$_{\rm s}$} photometry in 1\tdeg\ttimes1\tdeg\ portions of the
\WISE\ \scode{allsky\_4band\_p3as\_psd} catalog for a series of (RA,\,Dec)
that, together, fully cover \tsim12\tdeg\ttimes12\tdeg\ areas around both
NEP and SEP, generously encompassing the \JWST\ CVZs.\deleted{That grid of
(RA,\,Dec) was generated with a custom \scode{SuperMongo} macro,
\scode{getwctrs}.}  After retrieval, individual catalogs were concatenated
and sorted, retaining the unique sources.} \citep{WISE,NEOWISE} that
was centered on the NEP.  We analyzed the histogram of averaged 3.4 and
4.6\,\micron\ source magnitudes to select regions with the lowest densities
of bright sources.  In particular, we assigned an exponentially scaled
penalty to each source within the 10\tmin\ttimes10\tmin\ viewport, ranging
from 10$^{-7}$ for sources in the 1\,mag wide histogram bin centered at
\mAB\,=\,22\,mag to 10$^{10}$ for those in the bin at \mAB\,=\,4.0\,mag
(\ie penalties increase by a factor 10 for each magnitude increase in
brightness), and then summed the penalties over that viewport.
\emph{Very few} 10\tmin\ttimes10\tmin\ regions near the NEP at a Galactic
latitude of $b^{\rm II}$\,\tsim\,30\tdeg\ are devoid of stars brighter than
\mAB\,=\,16\,mag.  While there are candidate regions with 3 or fewer
\mAB\,\tsim\,16\,mag stars, \emph{only one} cluster of regions stood out in
that shifts of $+$ or $-$2\tmin\ are nearly as good as the very best one.
That most promising target region was centered on (RA,\,Dec)$_{\rm J2000}$
= (17:22:43,\,+65:49:36), and is indicated with small magenta circles to
the right of and below center in Figs.~\ref{fig:jwstCVZs_nir}\emph{a} and
\ref{fig:jwstCVZs_ebv}\emph{a}.  The former shows that this most promising
region appears indeed in a particularly dark spot in the sky.  The latter,
a map of Galactic reddening, shows that this lack of bright sources is not
due to higher Galactic foreground extinction: the extinction here averages
only $A_V$\,\lsim\,0.087\,mag or $A_K$\,\lsim\,0.015\,mag
(\textsl{E(B$-$V)}\,\lsim\,0.028\,mag), where we adopt the \citet{SF11}
recalibration of the \citet{SFD98} extinction map.

We then refined our moving viewport analysis, originally performed within
\code{SuperMongo}\footnote{\url{https://www.astro.princeton.edu/~rhl/sm/}},
by re-implementing it within a \code{Fortran77} program that allowed both
square and round viewports of arbitrary size, and allowed us to step through
a larger area on the sky, centered on either the NEP or the SEP, with
arbitrary step size.  This program also assigned source penalties to
individual sources directly, rather than to the sources cumulated in
1\,mag wide histogram bins.  We adopted a new normalization, such that a
source with a mean (3.4+4.6\micron) WISE source magnitude of 12.48
($m_{\rm Vega}$; corresponding to \mAB\,=\,15.5\,mag, given the offset of
3.02\,mag between the two systems at 4\,\micron\footnote{See the Explanatory
Supplement to the \WISE\ All-Sky Data Release Products (R.~Cutri \etal;
\url{http://wise2.ipac.caltech.edu/docs/release/allsky/expsup/sec4_4h.html\#conv2ab}, as updated August 7, 2017).})
results in a source penalty of exactly 1.  Hence, the source penalties
cumulated within a given viewport notionally add up to the equivalent number
of deeply saturating (\mAB\,=\,15.5\,mag) sources within that viewport.
Last, the new code was generalized to allow different weighting of 3.4 and
4.6\,\micron\ source magnitudes (3.4\micron\ only,
(3.4+4.6)/2\,\tsimeq\,4\,\micron, (2\ttimes3.4+4.6)/3, or 4.6\micron\ only).

The results for round viewports with a diameter of 14\tmin\ (our desired
\JWST\ survey field size; see \S\,2.2) stepped through the
12\tdeg\ttimes12\tdeg\ area centered on the NEP in 20$''$ steps is rendered
in Fig.~\ref{fig:jwstCVZs_srcpen}\emph{a} and, scaled identically, for the
SEP in Fig.~\ref{fig:jwstCVZs_srcpen}\emph{b}.  As in our preliminary
analysis, we find very few 14\tmin\ diameter regions that are devoid of
sources brighter than \mAB\,=\,15.5\,mag.  Nonetheless, within \JWST's
northern CVZ, a few small regions and clusters of such regions appear as
dark blue shades in Fig.~\ref{fig:jwstCVZs_srcpen}\emph{a}, the very best
cluster of which is still centered on (RA,\,Dec)$_{\rm J2000}$ =
(17:22:43,\,+65:49:36). We did \emph{not} identify any similarly clean
regions within the southern CVZ.

While \JWST's southern CVZ will offer great opportunities for time-domain
science in the LMC, we will discard it from further consideration for the
purpose of any deep extragalactic field to survey the distant universe.

\subsection{Point Sources versus Extended Sources}

Point sources at a given magnitude will have a larger impact on detector
persistence than more extended sources (galaxies).  For sources of the same
magnitude, the brightest pixels in the latter will be several magnitudes
fainter than in stellar images.  We therefore verified in a 2MASS
\citep{2MASS} \textsl{JHKs} composite image with an angular resolution of
\tsim2$''$ (FWHM) that most of the \tsim4\,\micron-bright \WISE\ sources
(nominal FWHM\,\tsim\,6\farcs1--6\farcs4) within the best cluster of clean
14\tmin\ diameter regions are indeed Galactic stars, and proved that the
best region selected is indeed devoid of bright \emph{red} stars.

\vspace*{4pt}

\begin{figure*}[ht]
\centerline{
  \includegraphics[width=\txw]{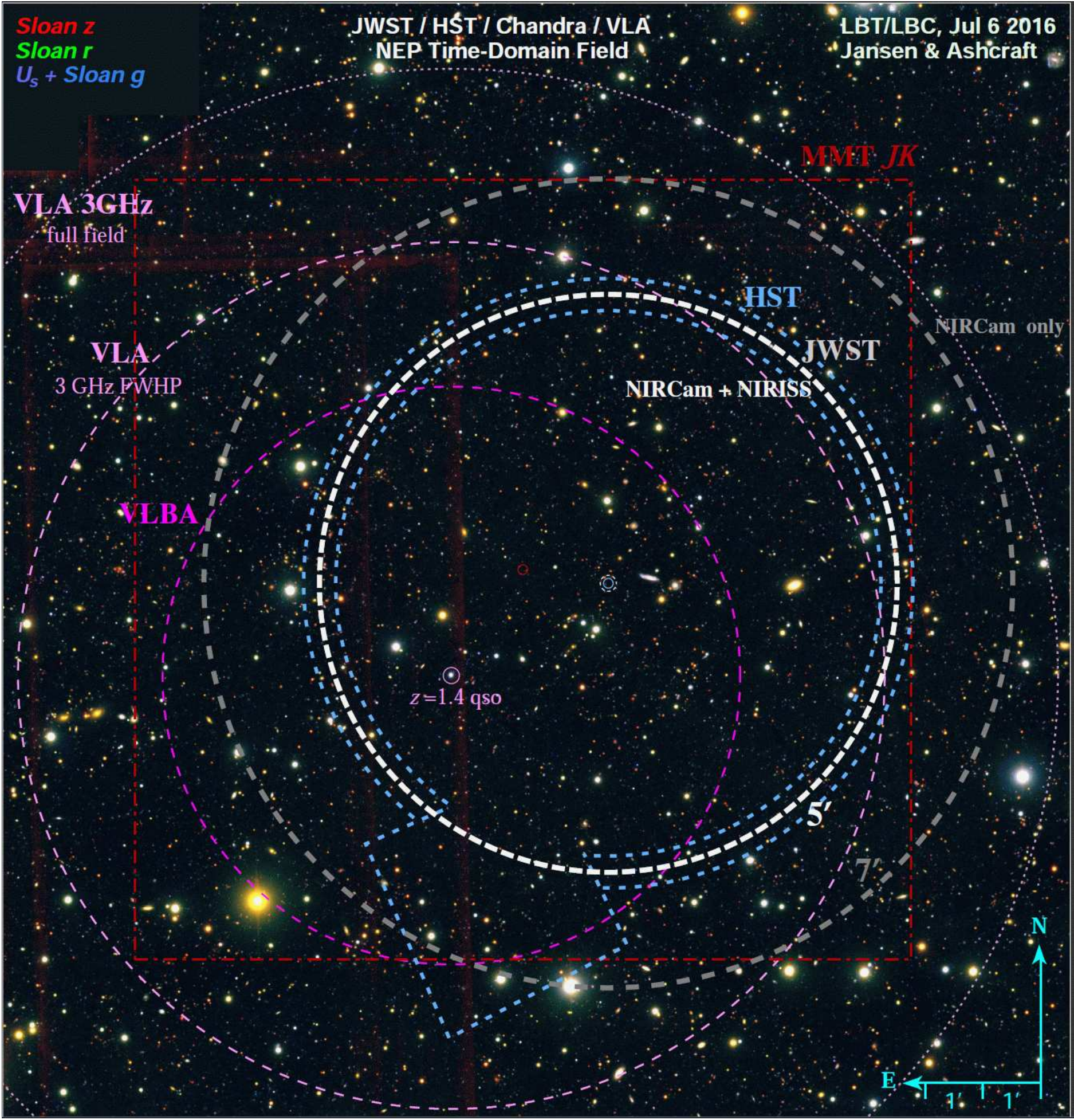}
}
\caption{\noindent\small
A 19\tmin\ttimes20\tmin\ \textsl{Ugrz} color image covering the \JWST\
Time-Domain Field (TDF) and its immediate surroundings, obtained by Jansen \&
Ashcraft using the 2\ttimes8.4\,m Large Binocular Telescope (LBT). The depth
in each of \textsl{U}, \textsl{g}, and \textsl{r} exceeds \mAB\,=\,26\,mag,
while \textsl{z} is only slightly shallower. There are no bright stars in
the NIRCam field (gray dashed circle), and the whole area has
\EBV$\lesssim$\,0.028\,mag.  The red square indicates extant MMT/MMIRS
\textsl{J} and \textsl{K} coverage (PI: C.~Willmer), while the pale blue
contour approximates our \HST\ WFC3/UVIS \textsl{F275W} and ACS/WFC
\textsl{F435W} and \textsl{F606W} coverage (PI: R.~Jansen;
\citealt{AAShst15278}). The pink dashed and outer dotted circles indicate
the half-power and full extent of our VLA 3\,GHz field (PI: R.~Windhorst);
the magenta dashed circle corresponds to the half-power VLBA coverage at
5\,GHz (PI: W.~Brisken; up to \tsim800 VLA-detected sources will be
followed up at high angular resolution with the VLBA). The VLA/VLBA pointing
center is on VCS5 \citep{VCS5} J172314.1+654746, a \tsim0.2\,Jy
flat-spectrum quasar at $z$\,=\,1.4429 that remains unresolved even by the
VLBA.  Scheduled \Chandra/ACIS\,I X-ray observations (PI: W.~Maksym) target
the entire area with \JWST\ and VLA coverage. This is the only clean survey
region in the sky with a perfect point source radio calibrator where \JWST\ 
can get NIRCam 1--5\,\micron\ imaging to 29\,mag \emph{and} overlapping
NIRISS 1.8--2.2\,\micron\ spectra to 28\,mag at \emph{any} time of the year:
the ideal time-domain field. The brightest yellow star prevents the \JWST\ 
field center from precisely coinciding with the radio field center.
\label{fig:jwstNEPtds_lbc}}
\end{figure*}

Moreover, in order to verify the suitability of this field to significantly
fainter limits, in July 2016 we secured deep \textsl{Ugrz} observations
with the Large Binocular Cameras (LBCs) on the 2\ttimes8.4\,m Large Binocular
Telescope (LBT) atop Mt.Graham, Arizona
\citep[][2019a \protect{[}in prep.\protect{]}]{AASlbtlbc2}.
Fig.~\ref{fig:jwstNEPtds_lbc} shows a color composite of a portion of the
full LBT/LBC images that covers the \JWST\ NEP TDF and its immediate
surroundings.  The individual \textsl{Ugrz} mosaics for that color composite
were produced following \citet{Ashcraftetal18}.

\vspace*{4pt}

The faintest sources discernable with white to orange hues
have \mAB\,\tsim\,26.0--26.5\,mag.  The effective resolution in this color
composite is \tsim0\farcs95 (FWHM).  There are no bright stars within the
field that NIRCam would cover in an implementation of \JWST\ observations
similar to those outlined in \S~4 (indicated by a gray dashed circle with a
radius of 7$'$), and there is no hint of Galactic cirrus, nor of filaments
or patches of dust.
In fact, the image shows a wide variety and large number of faint background
galaxies, including some groups and distant clusters of galaxies (orange
hues), along with a smattering of faint stars, as one would expect for a
field at intermediate Galactic latitude ($b^{\rm II}$\,\tsimeq\,+33.6\tdeg). 
For a detailed description of these LBT/LBC data, we refer the reader to
Jansen \etal\ 2019a (in prep.).

%%%%%%%%%%%%%%%%%%%%%%%%%%%%%%%%%  FIGURE 4  %%%%%%%%%%%%%%%%%%%%%%%%%%%%%%%%
%\placefigure{n}
\noindent
%%%%%%%%%%%%%%%%%%%%%%%%%%%%%%%%%%%%%%%%%%%%%%%%%%%%%%%%%%%%%%%%%%%%%%%%%%%%%

%%%%%%%%%%%%%%%%%%%%%%%%%%%%%%%%%  FIGURE 5  %%%%%%%%%%%%%%%%%%%%%%%%%%%%%%%%
%\placefigure{n}
\noindent\begin{figure*}[ht]
\centerline{
  \includegraphics[width=0.49\txw]{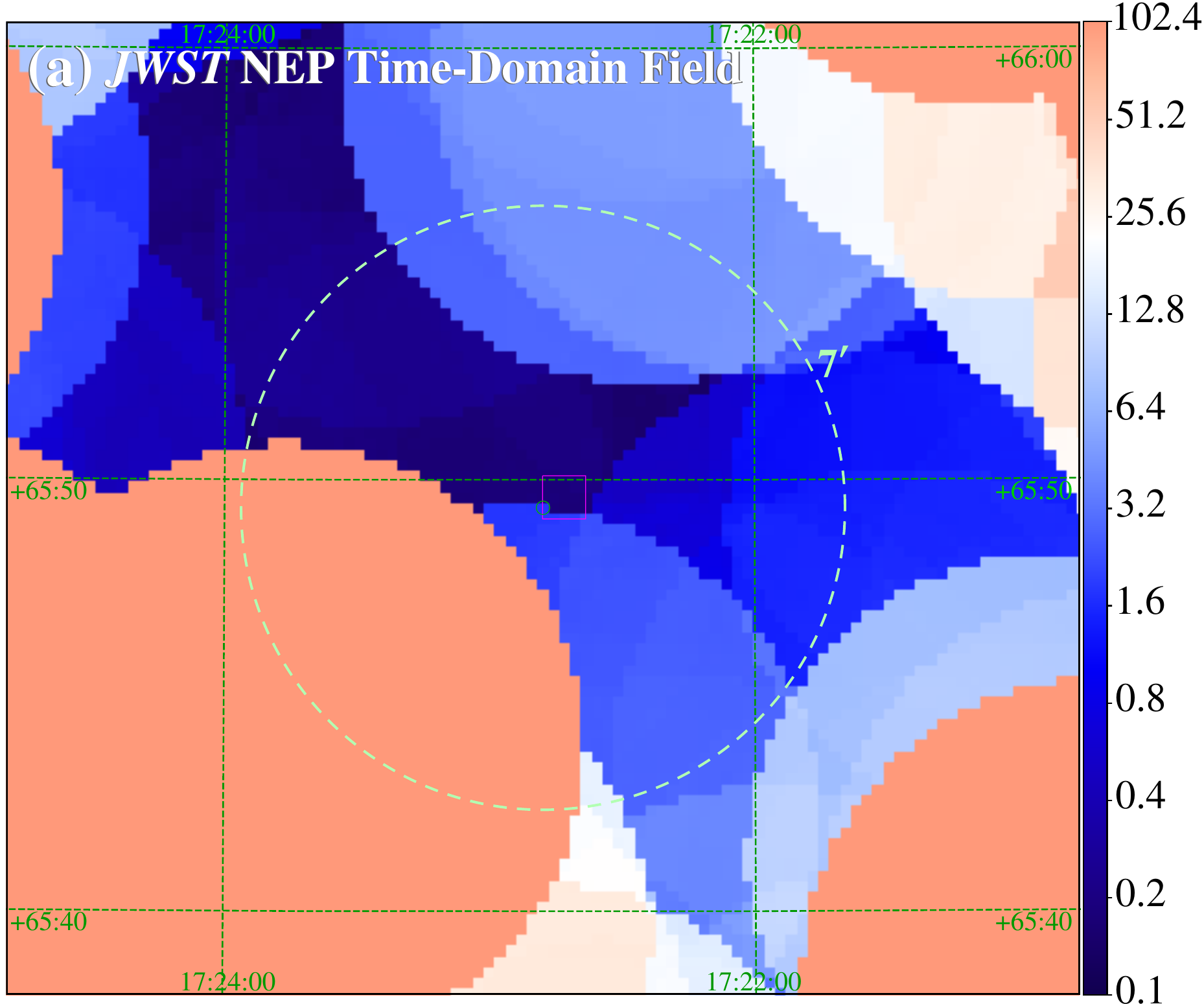}
  \hspace*{1mm}
  \includegraphics[width=0.49\txw]{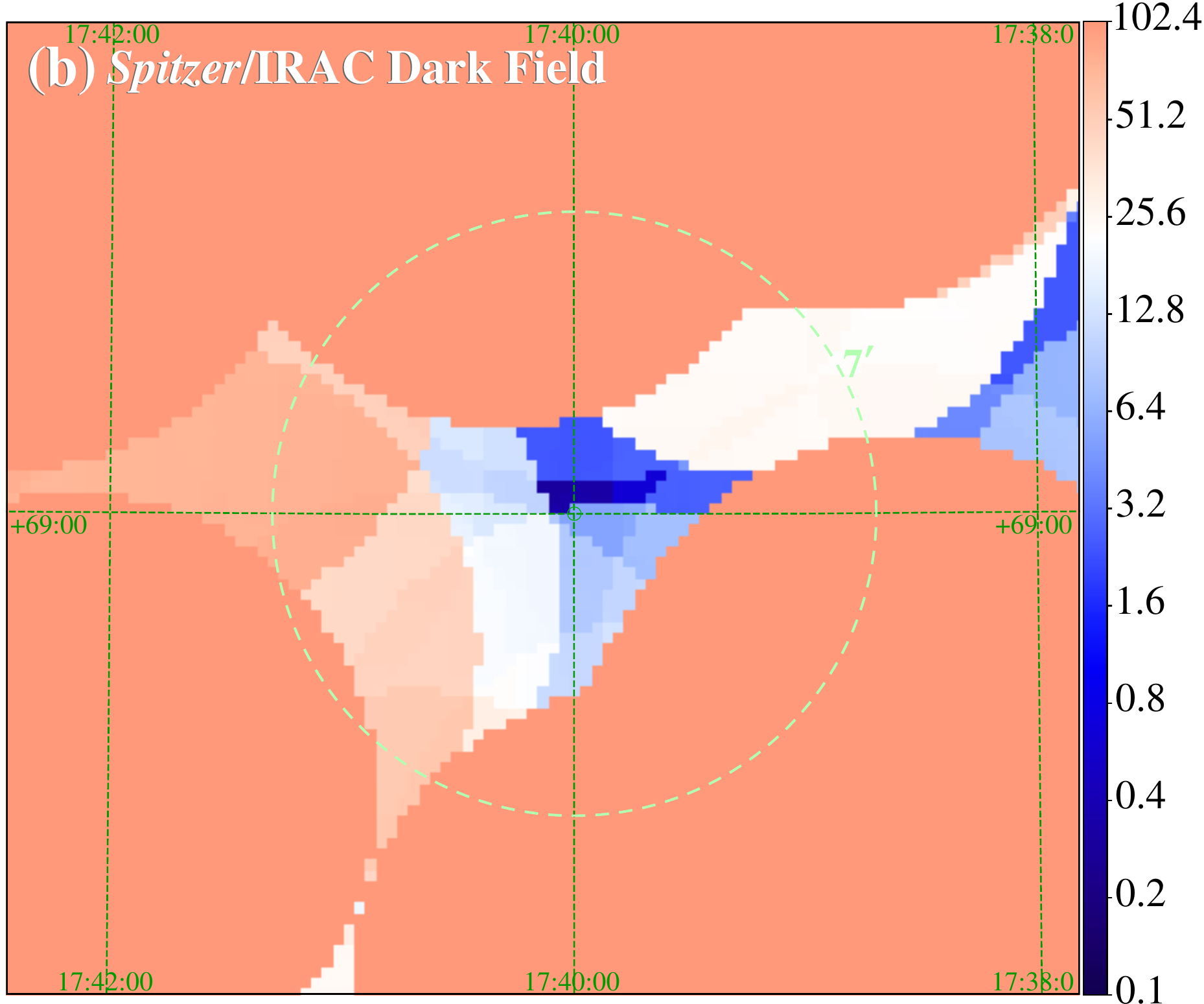}
}
\caption{Comparison of 18\tmin\ttimes18\tmin\ maps of the bright source
penalties of Fig.~\ref{fig:jwstCVZs_srcpen}\emph{a} multiplied by the
Galactic foreground reddening of Fig.~\ref{fig:jwstCVZs_ebv}\emph{a} for
[\emph{a}] the \JWST\ NEP TDF, and [\emph{b}] the \Spitzer/IRAC Dark Field
(IDF).  For a circular viewport with a diameter of 14\tmin, a step size of
15$''$, and source penalties that decrease by a factor 10 per unit
magnitude, the value at the center of the best survey field for \JWST\ 
(indicated by the magenta square) is \tsim0.18, compared to \tsim0.30 near
the center of the IDF.  While that difference is less than a factor 2, the
most striking feature of the comparison is that there is a much wider area
around the NEP TDF that is nearly as good as the NEP TDF proper, \ie there
is room for a survey that is much wider than can be accommodated at or near
the IDF without running into bright stars.  In fact, the dashed circle in
panel (\emph{a}) is shifted slightly east and south with respect to the
very best field center, in order to accommodate \HST\ guide star solutions
at multiple roll angles (see \S\,3.2.2).\label{fig:jwstTDFvsIDF1}}
\end{figure*}
%%%%%%%%%%%%%%%%%%%%%%%%%%%%%%%%%%%%%%%%%%%%%%%%%%%%%%%%%%%%%%%%%%%%%%%%%%%%%

\vspace*{-24pt}
\section{Discussion}

\subsection{The best continuously accessible survey field for
	{\itshape JWST}}

The very best field selected from our analysis of \tsim4\,\micron\ source
penalties in the \JWST's northern CVZ has central coordinates of
(RA,\,Dec)$_{\rm J2000}$ = (17:22:43.12, +65:49:36.0). 
Fig.~\ref{fig:jwstTDFvsIDF1}\emph{a} shows a more detailed map of these
source penalties weighted by the inverse of the Galactic foreground
reddening \EBV\ (\ie weighted towards the regions of lowest extinction).
There are adjacent good fields \tsim1--2\tmin\ toward the north, east, and
northwest, as well as a cluster of fields \tsim8\tmin\ to the northeast,
that are almost as good as this very best field in terms of both source
penalty and Galactic foreground extinction.  That makes this field ideal
for not just a \tsim14\tmin\ diameter \JWST\ time-domain survey, but also
for future extensions to a deep or medium-deep extragalactic survey that
covers a wider area.

%*% Note (prompted by referee):  JWST backgrounds at @4.4mu:
%*% Verified for NEP TDF originally dd. Dec 21-22 2016 using a demo version
%*% of the JWST ETC; verified & still correct dd. Sep 23 2018 using v1.2.2
%*% of the JWST ETC (https://jwst.etc.stsci.edu/):
%*%            darkest background   background for 180deg flip   bg.max
%*%   NEP TDF        0.23 MJy/sr           0.24 MJy/sr            ~0.26
%*%   GOODS-N/HDF-N  0.20                  0.27
%*%   GOODS-S/HUDF   0.19                  0.31
%*%   COSMOS         0.23                  0.53!!

It may be of interest to compare \tsim4\,\micron\ backgrounds and source
penalties within the NEP TDF with those in established deep fields like the
CANDELS \citep{CANDELS1,CANDELS2} GOODS-N (including the HDF-N), GOODS-S
(including the HUDF), and COSMOS fields.  While not continuously accessible
to \JWST, the best possible backgrounds at 4.4\,\micron\ in the middle of
their visibility windows are \tsim0.20, 0.19, and 0.23\,MJy/sr,
respectively, as compared to \tsim0.23\,MJy/sr for the NEP TDF\,$^2$.
  %
%*% Refer to footnote 2 on page 4 instead of repeating this a third time:
%*%\footnote{As reported by the \JWST\ ETC \citep{JWSTetc} v1.2 available
%*%at \url{https://jwst.etc.stsci.edu/}.}
  %
Backgrounds corresponding to dates of observation that allow a 180\tdeg\ 
flip in orientation are higher (\tsim0.27, 0.31, and 0.53\,MJy/sr), while
there is no restriction on date of observation for the \JWST\ NEP TDF, where
backgrounds are typically \tsim0.24\,MJy/sr and never higher than
\tsim0.26\,MJy/sr.
  %
%*% Source penalty values cumulated within 14' diameter circular viewports:
%*%        best with any HST coverage:   best with deep HST coverage:
%*% NEP TDF         6.4799                   6.4799
%*% GOODS-N/HDF-N  10.4777 (SE of CANDELS)  11.7986 (immediately NW of HDF-N)
%*% GOODS-S/HUDF    5.6161 (in CANDELS, SE of HUDF) 5.8306 (NW edge of HUDF)
  %
Source penalty values are very similar to or even slightly lower in the
GOODS-S/HUDF area than in the \JWST\ NEP TDF, with a similarly large choice
in placement of the field center.  In fact, \JWST\ could center a deep
14\tmin\ diameter survey that would be devoid of 4\,\micron-bright point
sources almost anywhere within the deep portion of the fiducial CANDELS
GOODS-S footprint as well as within a portion of the WFC3/ERS field
\citep{Windhorstetal11}. The CANDELS GOODS-N footprint contains one and is
significantly encroached by four relatively bright sources, leaving only a
highly constrained area free of such sources that is ideal for a deep
14\tmin\ diameter \JWST\ survey.  In that very best area, source penalty
values are \lsim2\ttimes\ higher than in the \JWST\ NEP TDF.

A \emph{time-domain} field must be accessible 365 days per year, however,
which implies a location within a CVZ.  In the following section we will
discuss considerations that might favor particular center coordinates that
are offset with respect to the best field center for \JWST.

\subsection{Other considerations}

\subsubsection{VLA Radio Interferometric Observations}

Deep Karl G.~Jansky Very Large Array (VLA) radio observations can be
obtained most efficiently if a suitable, unresolved phase calibration
source is located exactly in the center of the synthesized beam.  The best
14\tmin\ diameter field for \JWST\ includes a
$m_{z{\rm ,AB}}$\,\tsim\,16.9\,mag \citep[SDSS;][]{SDSS} flat-spectrum
quasar at $z$ = 1.4429 \citep{SDSSqsozs} with a NVSS \citep{NVSS} flux
density of 239.4\tpm7.2\,mJy at 1.4\,GHz, and \citep[VCS5;][]{VCS5} VLBA
flux densities of \tsim230\,mJy at 2.3\,GHz and \tsim140\,mJy at 8.6\,GHz,
respectively.  In the VCS5 catalog, this quasar is classified as a suitable
phase calibrator that remains unresolved\footnote{See, \eg
\url{https://gemini.gsfc.nasa.gov/results/vcs/vcs5/vcs5_cat.html}\,.}
at VLBA resolution, and has coordinates (RA,\,Dec)$_{\rm J2000}$ =
(17:23:14.1381, +65:47:46.178), \ie \tsim2\farcm7 to the southeast of the
very best field center for {\JWST}\/.  VLA 3\,GHz ($\lambda$\,=\,10\,cm;
PI: R.~Windhorst) and VLBA 5\,GHz (6\,cm; PI: W.~Brisken) observations to
\muJy\ sensitivities, centered on these coordinates, started in November
2017 and are ongoing. Fig.~\ref{fig:jwstNEPtds_lbc} indicates the location
of the phase calibrator, and shows the footprints of these radio
observations as pink dotted and dashed, and magenta dashed circles,
respectively.

The ability to study variable AGN and cataclysmic events in both radio and
near- to mid-infrared at similar resolutions, in order to trace the
co-evolution of supermassive blackholes and their host-spheroids over
cosmic time, was deemed sufficiently compelling to consider a move of the
\JWST\ field. The field cannot be moved over all the way to be centered on
the $z$\,=\,1.4429 quasar, however, since that would cause a bright red
star capable of saturating the NIRCam detectors to enter the field of view,
resulting in unacceptable persistence.  The IDS \JWST\ GTO observations of
R.~Windhorst as specified and implemented in \code{APT}\footnote{See
\emph{JWST Astronomer's Proposal Tool, APT}, JWST User Documentation, 2018,
STScI (Baltimore, MD);
\url{https://jwst-docs.stsci.edu/display/JPP/JWST+Astronomers+Proposal+Tool\%2C+APT}. \scode{APT} is available from \url{http://apt.stsci.edu/}\/.}
versions prior to 25.2.1 (2017 July 6) were therefore moved to a compromise
field center of (RA,\,Dec)$_{\rm J2000}$ = (17:23:02.55, +65:49:36.0), such
that the \JWST/NIRCam+NIRISS survey footprint would fall entirely within
the VLA 3\,GHz half-power beam width, and most of the NIRISS survey
footprint would be encompassed by the VLBA 5\,GHz half-power beam width.
It is at this compromise position that ancillary ground-based near-infrared
imaging in \textsl{J} and \textsl{K} were secured in 2017 with MMT/MMIRS
(PI: C.~Willmer) as indicated by the dark red dot-dashed square in
Fig.~\ref{fig:jwstNEPtds_lbc}.  Future papers will describe and analyse
these and other ancillary data (mentioned below) in detail.

\subsubsection{Hubble Space Telescope observations}

UV--visible imaging with \HST\ of the central 5\tmin\ radius portion of
the \JWST\ NEP TDF, plus an extension toward the south-southeast to
$r$\,\tsim\,7\tmin, using WFC3/UVIS (F275W) and ACS/WFC (F435W and F606W)
was approved for Cycle~25 (GO-15278; PI: R.~Jansen).
Simultaneous \HST\ guide star availability for a pattern with 8 effective
field centers and 9 distinct orientations, each highly constrained to be
able to make use of \HST's scarce near-CVZ opportunities for this field
\citep[see][]{AAShst15278} proved a problem at the VLA+\JWST\ compromise
field center.  By shifting the field center 90$''$ west (\ie back in the
direction toward the objectively selected best field center) and 14$''$
south, a viable solution was found, however, for (RA,\,Dec)$_{\rm J2000}$ = 
(17:22:47.896, +65:47:21.54), as indicated in Fig.~\ref{fig:jwstNEPtds_lbc}
by a small pale blue circle.  This field center was subsequently adopted
(small dashed white circle) for the IDS \JWST\ GTO observations of
R.~Windhorst (indicated by the white and grey dashed circles with radii of
5$'$ and 7$'$, respectively) and H.~Hammel as implemented in \code{APT}
version 25.4.2, and made public by the \JWST\ Project on Feb. 6, 2018,
making it the \emph{de~facto} coordinates of the \JWST\ NEP TDF.  Since
the VLA still has significant sensitivity beyond its half-power beam width,
all of the \JWST/NIRCam observations in the NEP TDF will still have
coverage at 3\,GHz.

\subsubsection{Other proposed, approved, and scheduled observations}

These \emph{de~facto} coordinates of the \JWST\ NEP TDF were also adopted
for proposed \JWST\ ERS observations in this field (MIRI+NIRCam imaging:
H.~Messias \etal; NIRISS wide-field slitless spectroscopy: S.~Malhotra
\etal), as well as for \Chandra/ACIS\,I X-ray imaging in progress (Cycle~19
program 19900666; PI: W.~Maksym) and approved (time-domain monitoring
Cycle~20+21 program 20900658; PI: W.~Maksym), for IRAM\,30m/NIKA2 mm-wave
observations in progress (PI: S.~Cohen), for approved and proposed
JCMT/SCUBA2 (PI: I.~Smail, M.~Im), LOFAR (PI: P.~Best), and 4.2\,m WHT/WEAVE
(PI: K.~Duncan) observations, for MMT/Binospec multi-object spectroscopy and
MMT/MMIRS \textsl{Y} (some \textsl{H}) near-IR imaging in hand (PI: 
C.~Willmer), and proposed \Spitzer/IRAC (PI: M.~Ashby) imaging, and SMA (PI:
G.~Fazio) sub-mm interferometry. The field and its wider surroundings have
also been targeted with Subaru/HSC in five filters
(\textsl{g,i,z,\,NB$_{816}$,\,NB$_{921}$}) as part of the Hawaii EROsita
Ecliptic-pole Survey (HEROES; PI: G.~Hasinger), and with the J-PAS
\citep{J-PAS} PathFinder for spectrophotometric narrow-band imaging (PI:
S.~Bonoli and R.~Dupke). Ground-based time-domain \textsl{ugriz}
observations with HiPERCAM on the 10\,m GTC (PI: V.~Dhillon) are also
proposed.
  %
%%%%%%%%%%%%%%%%%%%%%%%%%%%%%%%%%  FIGURE 6  %%%%%%%%%%%%%%%%%%%%%%%%%%%%%%%%
%\placefigure{n}
\noindent\begin{figure*}[ht]
\centerline{
  \includegraphics[width=0.48\txw]{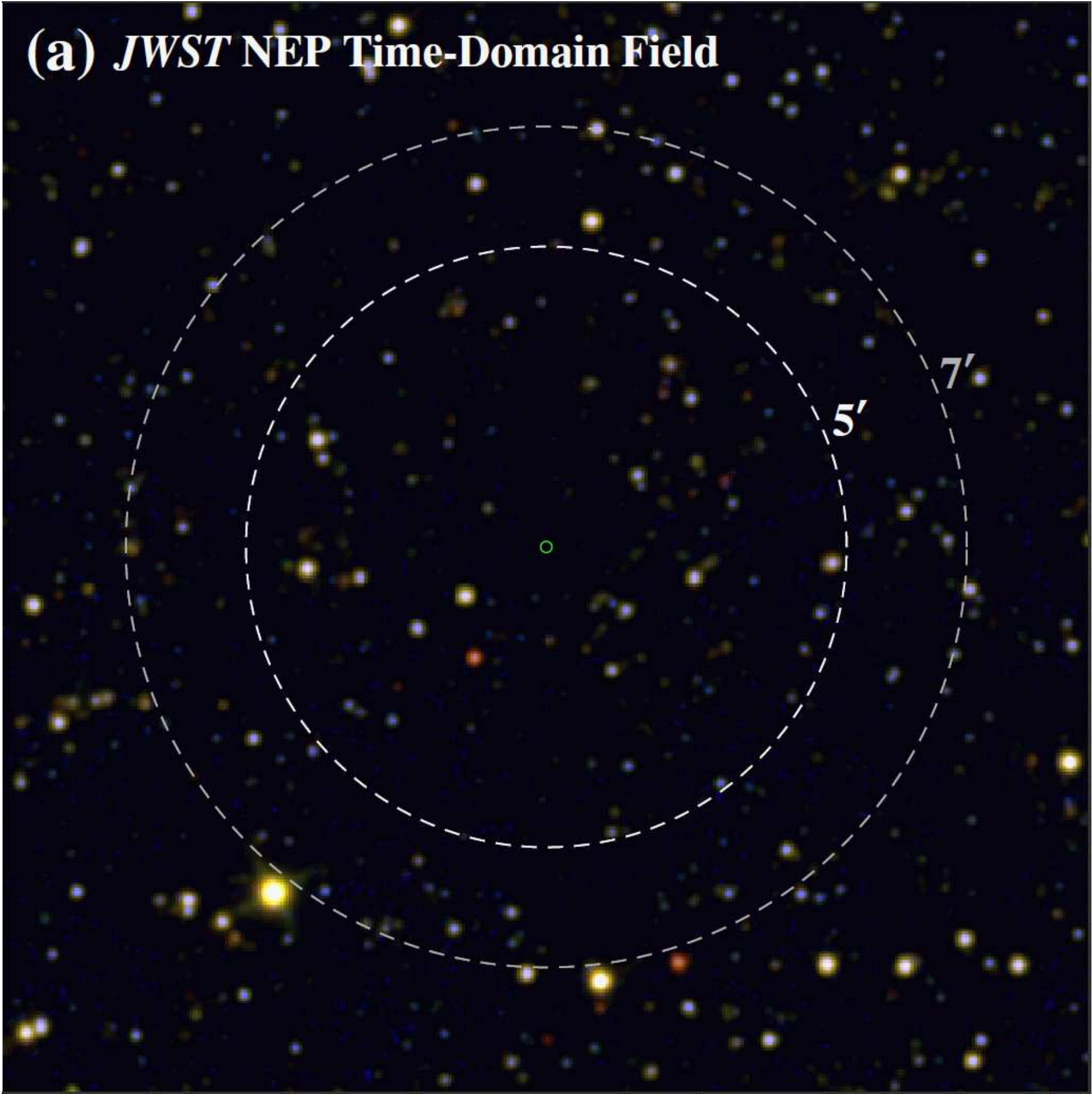}
  \hspace*{4mm}
  \includegraphics[width=0.48\txw]{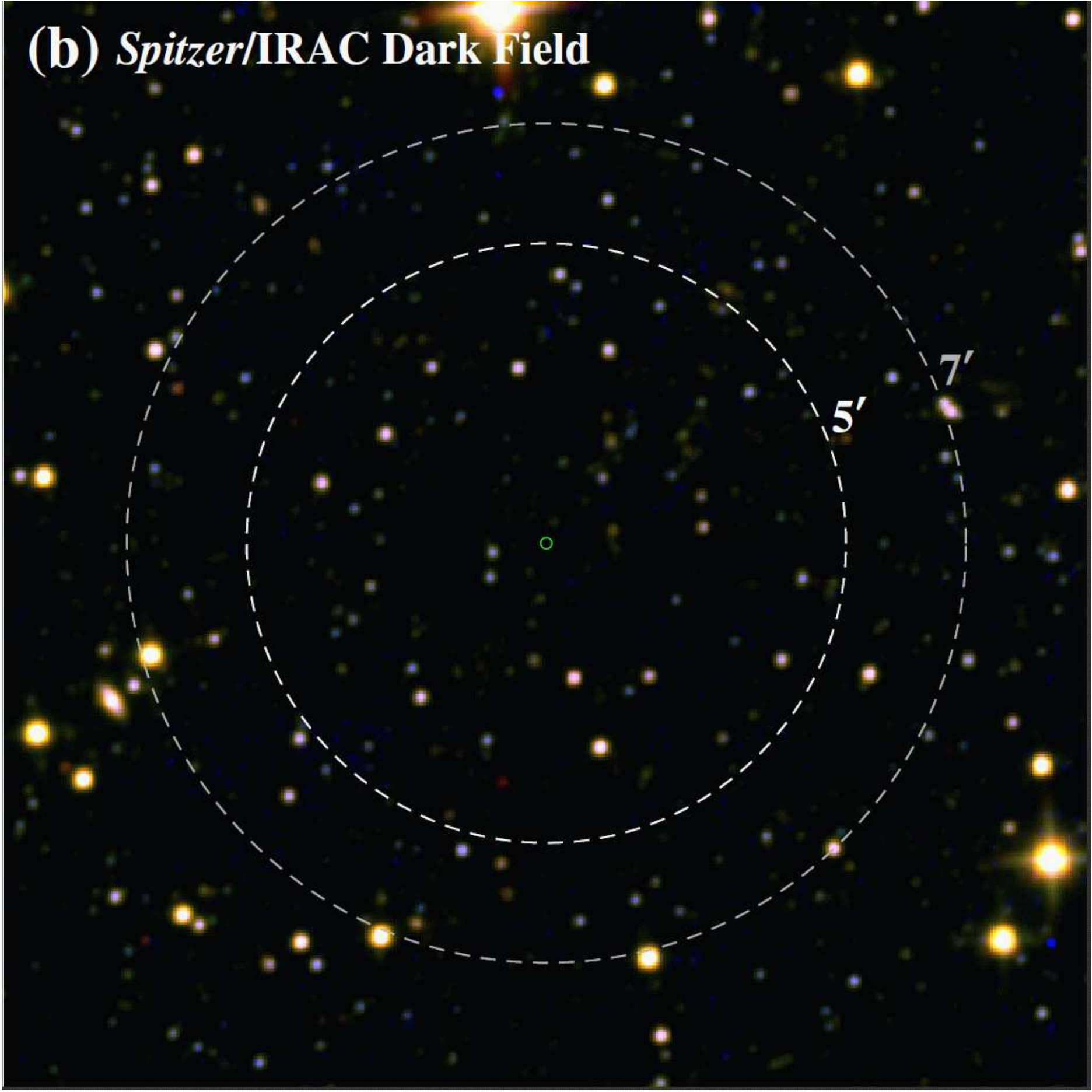}
}
\caption{Comparison of near-infrared color composites of [\emph{a}] the
selected \JWST\ NEP Time-Domain Field, and [\emph{b}] the \Spitzer/IRAC Dark
Field.  In each 18\tmin\ttimes18\tmin\ image, 2MASS \textsl{J}, \WISE\ 
3.4\,\micron, and \WISE\ 4.6\,\micron\ images are shown at \tsim6\farcs4
resolution (FWHM) in blue, green, and red hues.  Dashed white and grey
circles have radii of 5\tmin\ and 7\tmin, respectively.  While the central
portion of the IDF is nearly as free of bright sources as the NEP TDF, the
latter can be moved or expanded a few arcminutes up (north), and then left
(east) or right without including additional bright sources.\\
$\quad$%\linebreak
\label{fig:jwstTDFvsIDF2}}
\end{figure*}
%%%%%%%%%%%%%%%%%%%%%%%%%%%%%%%%%%%%%%%%%%%%%%%%%%%%%%%%%%%%%%%%%%%%%%%%%%%%%

%\section{Why not the IRAC Dark Field}
\subsection{A critical comparison of the {\itshape JWST} NEP TDF and the
	{\itshape Spitzer}\,/\,IRAC Dark Field}

Every two to three weeks throughout its mission, {\Spitzer/} IRAC
\citep{Spitzer/IRAC} observations were made of a field near the NEP and
centered at (RA,\,Dec)$_{\rm J2000}$ = (17:40:00, +69:00:00), the IRAC Dark
Field (IDF), for the purpose of dark and bias calibration of its detectors.
This field is also located well within the northern \JWST\ CVZ, and its
original selection also aimed to target an empty patch of sky, avoiding
bright stars and extended galaxies. Over time, this has resulted in the
deepest extant near-IR extragalactic survey field, with a \tsim15 year
baseline for time-domain studies in its shortest wavelength bandpasses.
The field was also augmented with deep observations at other wavelengths
from X-ray through far-infrared \citep{IDF}.  The clean portion of the
\tsim20\tmin\ diameter IDF to the depths reached by \deleted{\JWST\ and
considering the exquisite sensitivity of}\JWST/NIRCam is much smaller,
however, and of the order of no more than $r$\,\tsim\,7\tmin{}, as can be
seen in Fig.~\ref{fig:jwstTDFvsIDF2}\emph{b}.

\newpage

\vspace*{8pt}
Indeed, Fig.~\ref{fig:jwstCVZs_srcpen}\emph{a} shows that on scales of
14\tmin, the IDF does not stand out as an optimal field for \JWST.
Fig.~\ref{fig:jwstTDFvsIDF1}\emph{b} emphasizes that, while (weighted)
source penalties in 14\tmin\ diameter circular viewports are only slightly
higher in the IDF than in the very best area selected for the \JWST\ 
Time-Domain Field, the latter field can accommodate a deep \JWST\ survey
over a much wider area.  Figs.~\ref{fig:jwstTDFvsIDF2}\emph{a} and
\ref{fig:jwstTDFvsIDF2}\emph{b} compare the TDF and IDF in a near-infrared
color composite that directly shows the low densities of bright sources
within and encroachment of bright sources around the perifery of the IDF,
while the TDF has ample room to expand to the north, northeast, and west.
For reference, the brightest star encroaching the IDF toward the north has
$m_{K{\rm ,AB}}$ = 7.69\,mag, whereas the brightest star near the TDF
toward the southeast has $m_{K{\rm ,AB}}$ = 10.73\,mag.
Also when comparing the mean (median) \tsim2\,\micron\ brightness of
2MASS-detected stars within the cleanest central areas ($r$\,$\le$\,7\tmin)
of the IDF and TDF, we find that the TDF fares slightly better:
$\langle m_{K{\rm ,AB}}\rangle$ = 16.17 (16.65) for the IDF versus 16.65
(16.92) mag for the TDF.  The negative impact due to persistence effects
on deep \JWST\ surveys is therefore expected to be slightly reduced in the
TDF compared to the IDF.

%*% NB: from the WISE Source Catalog (RAJ's file "NEP_wiseSCs.dat"), which
%*%   lists 2MASS JHKs photometry in addition to WISE photometry:
%*%   * the brightest star near JWST NEP TDF is \mAB(\textsl{Ks}) =
%*%       m_{Vega}(Ks) + 1.85 = 8.877 + 1.85 = 10.73\,mag
%*%   * the brightest star near IRAC Dark Field is \mAB{\textsl{Ks}) =
%*%       m_{Vega}(Ks) + 1.85 = 5.838 + 1.85 = 7.69\,mag
%*%   * mean/median m_AB(Ks) of stars in the IDF is 16.170 / 16.65 mag
%*%   * mean/median m_AB(Ks) of stars in the TDF is 16.649 / 16.92 mag

\vspace*{6pt}

While well matched to the field of view of the \Spitzer/IRAC detectors, the
clean portion of the IDF is too small for our intended deep \JWST\ imaging
and slitless spectroscopic observations, which require a clean area of at
least 14\tmin\ diameter.

\subsection{The Promise of Parallel Observations}

Parallel science observations with \JWST\ were originally neither planned,
nor permitted.  In support of efficient on-orbit instrument calibration
after launch, ``parallel instrument calibrations'' were, however, to be
implemented \citep[\eg][]{JWST}.  In 2015, the \JWST\ Project and STScI
decided that parallel \emph{science} observations ought to be implemented
as well, in order to maximize the scientific return of {\JWST}\/. This is
especially important in view of the finite lifetime of \JWST\ ---as set by
its finite supply of fuel for station keeping and momentum control--- of 5 
(required), 10 (expected), or at most 14 (goal) years, which is much shorter
than that of \HST\ (28 years and counting).

During most of its operational lifetime, the depth and areal efficiency of
\HST\ grism spectroscopic surveys has added valuable low-resolution spectra
to many of the deepest \HST\ imaging surveys, which was essential to
measure the redshifts and characterize the properties of faint objects.
Prime examples of such \HST\ grism surveys are the Grism ACS Project for
Extragalactic Science \citep[GRAPES; e.g.,][]{GRAPES1,GRAPES2}, the ACS
Probing Evolution and Reionization Spectroscopically \citep[PEARS; 
e.g.,][]{PEARS3,PEARS1,PEARS2}, the WFC3 grism survey \citep[3DHST;
e.g.,][]{3DHST}, the WFC3 Infrared Spectroscopic Parallel survey \citep[WISP;
e.g.,][]{WISP}, the Grism Lens-Amplified Survey from Space \citep[GLASS;
e.g.,][]{GLASS1,GLASS2}, and the Faint Infrared Grism Survey \citep[FIGS;
e.g.,][]{FIGS1,FIGS2}.  We note that of these only WISP was an \HST\ parallel
survey.  The others targeted areas with pre-existing deep imaging, and
obtained grism spectroscopy with the primary instrument, while adding imaging
parallels.  Due to \HST's extraordinary longevity, grism spectroscopy and
direct imaging surveys could be completed in a largely separate manner. For
\JWST, with its much shorter expected life-time, direct imaging and slitless
grism spectra (NIRISS; especially for blind emission-line searches) or
multi-object slit-spectra (NIRSpec; targeted spectroscopy to the faintest
limits) must be done \emph{concurrently} to the largest extent possible.
Just like for \HST, we expect that \JWST\ NIRISS grism and NIRSpec
multi-object spectroscopy will constitute an essential complement to NIRCam
imaging observations to even fainter limits (\mAB\,\gsim\,27\,mag). Since we
anticipate that GO proposers will revisit the \JWST\ NEP Time-Domain Field
many times, the coordinates of even very faint objects will generally be
known in time to allow targeted spectroscopy and imaging in parallel.

\subsection{Estimated Source Counts and Number of Variable Objects in the
	{\itshape JWST} NEP Time-Domain Field}

%*% Chandra/ACIS-I: FOV = 16'x16' = 256 arcmin^2 for 4 ACIS-I chips
%*% VLA @3GHz: FOV (FWHP) has r ~ 0.5*45/nu[GHz] ~ 7.50' --> ~177 arcmin^2
%*%    (actual fov extends beyond the HP radius by a factor ~1.4x)
%*% VLBA @5GHz: r ~ 2/3 x VLA @3GHz ~ 5.0' --> 78.5 arcmin^2
%*% JWST NIRCam:        FOV has r ~ 7' --> 153.9 arcmin^2
%*% JWST NIRCam+NIRISS: FOV has r ~ 5' --> 78.5 arcmin^2
%*%   GTO1176: 4 spokes cover  71.83 arcmin^2 with NIRCam (IDS GTO)
%*%   GTO1176: 4 spokes cover  45.06 arcmin^2 with NIRISS (IDS GTO)
%*%   GTO+GO:  8 spokes cover 114.39 arcmin^2 with NIRCam (tentative)
%*%   GTO+GO:  8 spokes cover  59.65 arcmin^2 with NIRISS (tentative)
%*% Below we will estimate source counts for the area covered by the
%*% GTO1175 4-spoke program, and for the full 153.9 arcmin^2 corresponding
%*% to a circular area with a diameter of 14' (in common with pretty much
%*% all extant data radio and X-ray data).
%*%
%*% Source count slope at 3 GHz from Windhorst et al. (1985) and Hopkins
%*% et al. (2000) favors a value near -2.2; Condon et al. (2012) argue
%*% for a shallower slope at muJy levels near ~-1.7 . The devil is in the
%*% details of the various corrections, but here we will follow the more
%*% recent reference.

Deep \HST\ observations can inform our expectations for the near-IR source
counts within the \JWST\ NEP TDF.  To \mAB\,\tsim\,29\,mag, the HUDF
1.6\,\micron\ (\textsl{H}; as most representative of \JWST) counts show
3.5\ttimes10$^5$ objects per 0.5 mag interval per square degree (Fig.~12 of
\citealt{Windhorstetal11}). The 1.6\,\micron-counts over the
20\,\lsim\,\mAB\,\lsim\,29\,mag range show a sub-converging slope of
0.213\,mag\,dex$^{-1}$ \citep[see also][]{Driveretal16}.  Therefore, the
\emph{total} integrated galaxy counts to \mAB\,\lsim\,29\,mag are expected
to reach \tsim1.79\ttimes10$^6$ objects per deg$^2$.  The number of
Galactic stars at the NEP to the same flux limits is only
(0.4--0.9)\ttimes10$^4$ per deg$^2$ \citep{Ryanetal11,RyanReid16}, which is
still a \tsim5--10\ttimes\ larger density of stars than seen in the
\HST\ surveys summarized by \citet{Windhorstetal11} at high Galactic
latitude.

Similarly, given typical faint radio source counts \citep{Condonetal12,
Vernstrometal14, Windhorstetal85, Windhorstetal93, Hopkinsetal00}, we
expect that a VLA survey that reaches a depth of 5\,\muJy\ (5$\sigma$) at
3\,GHz would detect a \mbox{\emph{total}} of \tsim2.2\ttimes10$^4$ sources
per square degree. This calculation assumes that the normalized differential
source count slope continues to be as steep as $-$1.7 at \muJy\ levels
\citep{Condonetal12}, and a spectral index between 3 and 1.4\,GHz for these
sources of about 0.4, following the trends seen at somewhat brighter levels
between 1.4, 5.0, and 8.4\,GHz \citep{Windhorstetal93}.
The models of \citet{Hopkinsetal00} indicate that somewhat more than half
of the radio sources seen at \muJy\ levels will turn out to be starburst
galaxies or vigorously star forming regions within galaxies, while the
remainder are caused by weak AGN.  The latter would appear as unresolved
radio sources. The exact ratio of starbursts and AGN will of course await
VLBA observations at \muJy\ levels and the \HST\ + \JWST\ images and grism
spectra.

Both VLA 3\,GHz and \Chandra\ X-ray coverage will be available for the
\JWST\ NEP TDF, as well as significant partial coverage with \JWST/NIRISS
grism and \HST\ UV--Visible imaging observations.  If we consider the
\tsim153.9\,arcmin$^2$ and \tsim71.8\,arcmin$^2$ areas of the full
14\tmin\ diameter \JWST\ NEP TDF and of the footprint that will be sampled
in our initial \JWST\ NIRCam GTO observations (\S\,4), then we estimate a
total number of faint galaxies (stars) detected to \mAB\,\tsim\,29\,mag of
\tsim7.65\ttimes10$^4$ (170--385) and \tsim3.57\ttimes10$^4$ (80--180),
respectively.  In the same \JWST\ NEP TDF areas, we expect to detect a
total of \tsim940 and \tsim440 faint 3\,GHz radio sources to 5\,{\muJy},
almost half of which will be weak radio-selected AGN.

In an optical search for weak AGN, \citet{Sarajedinietal11} and
\citet{Cohenetal06} monitored faint galaxies at $z$\,\tsimeq\,0.5--4 with
\HST\ WFPC2 or ACS/WFC at visible (restframe UV) wavelengths on timescales
of weeks to months (\ie about a week to a month in the restframe at the
median redshift of $z$\,\tsim\,2 of the sample). These
studies find that on such timescales typically \tsim1\% of the faint
optically selected galaxies at a given redshift shows significant evidence
for variability in their cores (45 out of 4644 galaxies in the HUDF), about
half of which are associated with either faint X-ray sources or mid-IR
power-law emission \citep{Sarajedinietal11}.  That paper reports, moreover,
that a quarter of X-ray selected AGN are optical variables, and that this
percentage increases with decreasing hardness ratio of the X-ray emission.
At radio wavelengths, \citet{OfekFrail11} found in a NVSS--FIRST comparison
that 0.1\% of the unconfused FIRST sources in a mJy sample were variable.
\citep{OortWindhorst85} studied variability in a deeper sample of sub-mJy
radio sources by comparing deep Westerbork and VLA maps of the same region
in the Lynx\,2 field on time-scales of a month to a year.  They found that
\lsim2\% of the sub-mJy sources showed variability on both timescales.

Conservatively, within the full \JWST\ NEP TDF and within the GTO-sampled
portion thereof, we therefore estimate \gsim730 and \gsim340 galaxies
detected to \mAB\,\tsim\,29\,mag at rest-frame optical wavelengths to vary
in repeat observations with \JWST/NIRCam on time-scales of weeks to
years. We expect at least \tsim0.5\% of the field galaxies in the same
areas to be detected at \muJy\ levels with the VLA, and confirmed with VLBA
observations to be AGN (\ie \gsim380 and \gsim175 in total).  Starburst
dominated \muJy\ radio sources should be easily recognized from the 8-band
\JWST\ photometry and grism spectra, and are expected to appear resolved in
the VLBA observations, at least partially resolved in the
shorter-wavelength near-IR \JWST\ filters, and mostly non-varying at radio
and rest-frame visible--near-IR wavelengths, although a subset of
dust-obscured ULIRGs may be found to vary at mid--far-IR wavelengths on
time-scales of months to years due to their extremely high SN rates
\citep{Yanetal18}.%(submitted to ApJL)

Since any variable emission from weak AGN comes from different regions
around the accretion disk at radio, near--mid-IR, optical, and X-ray
wavelengths, the exact fraction of variable objects at each of these
wavelengths depends on the relative depth and time-sampling of each of
these data sets and on the physical timescales involved in each of these
wavelength regimes.  With a sufficient number of epochs and sufficient
depth from radio to X-rays, we hope that the \JWST\ NEP TDF will provide
the database to begin to address these questions, including how large a
fraction of weak AGN (variable or not) are seen in common between the
radio, near--mid-IR, optical, and X-ray studies.

%%%%%%%%%%%%%%%%%%%%%%%%%%%%%%%%%  FIGURE n  %%%%%%%%%%%%%%%%%%%%%%%%%%%%%%%%
%\placefigure{n}
\noindent\begin{figure*}[ht]
\centerline{
  \includegraphics[height=0.245\txw]{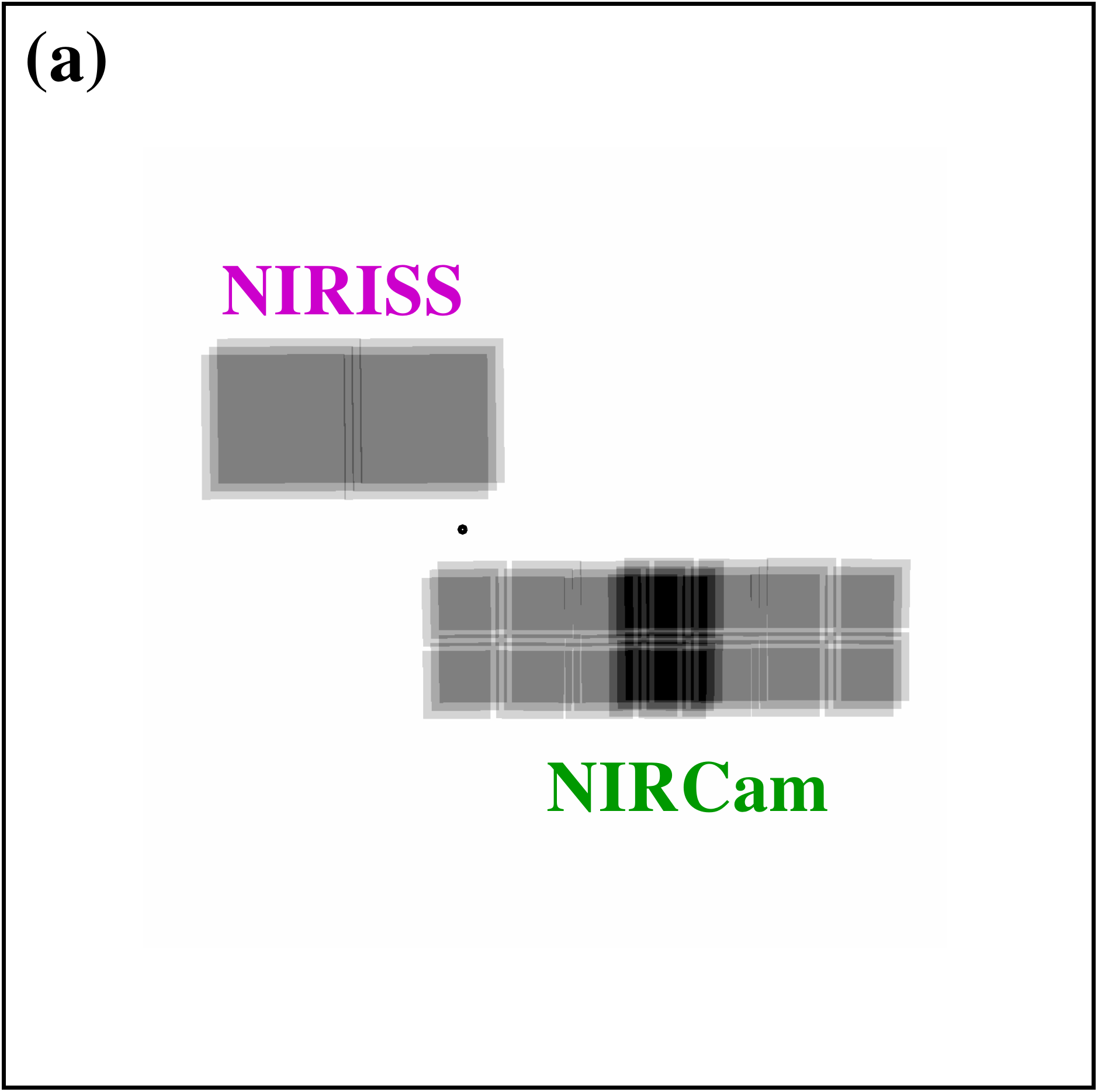}
  \includegraphics[height=0.245\txw]{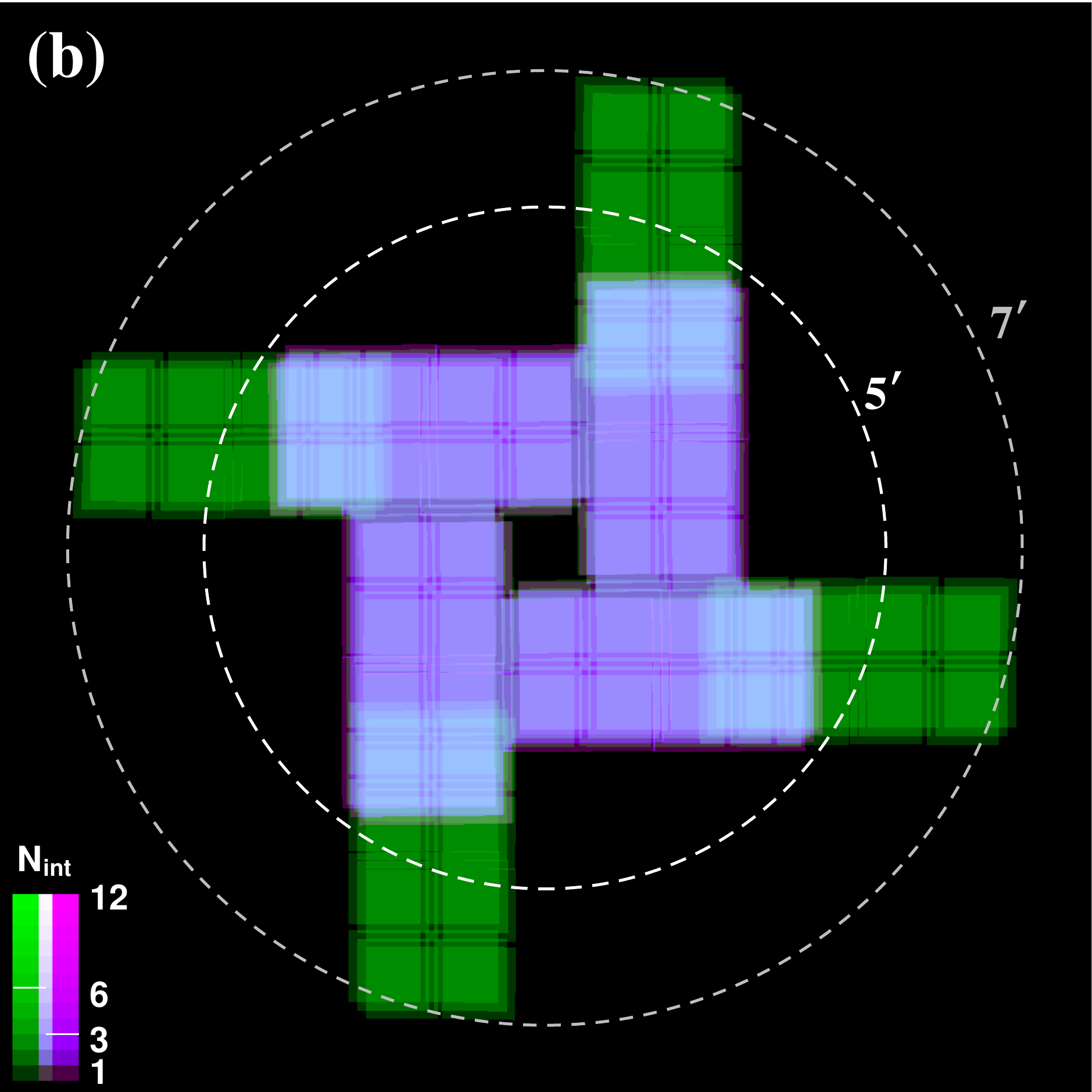}
  \includegraphics[height=0.245\txw]{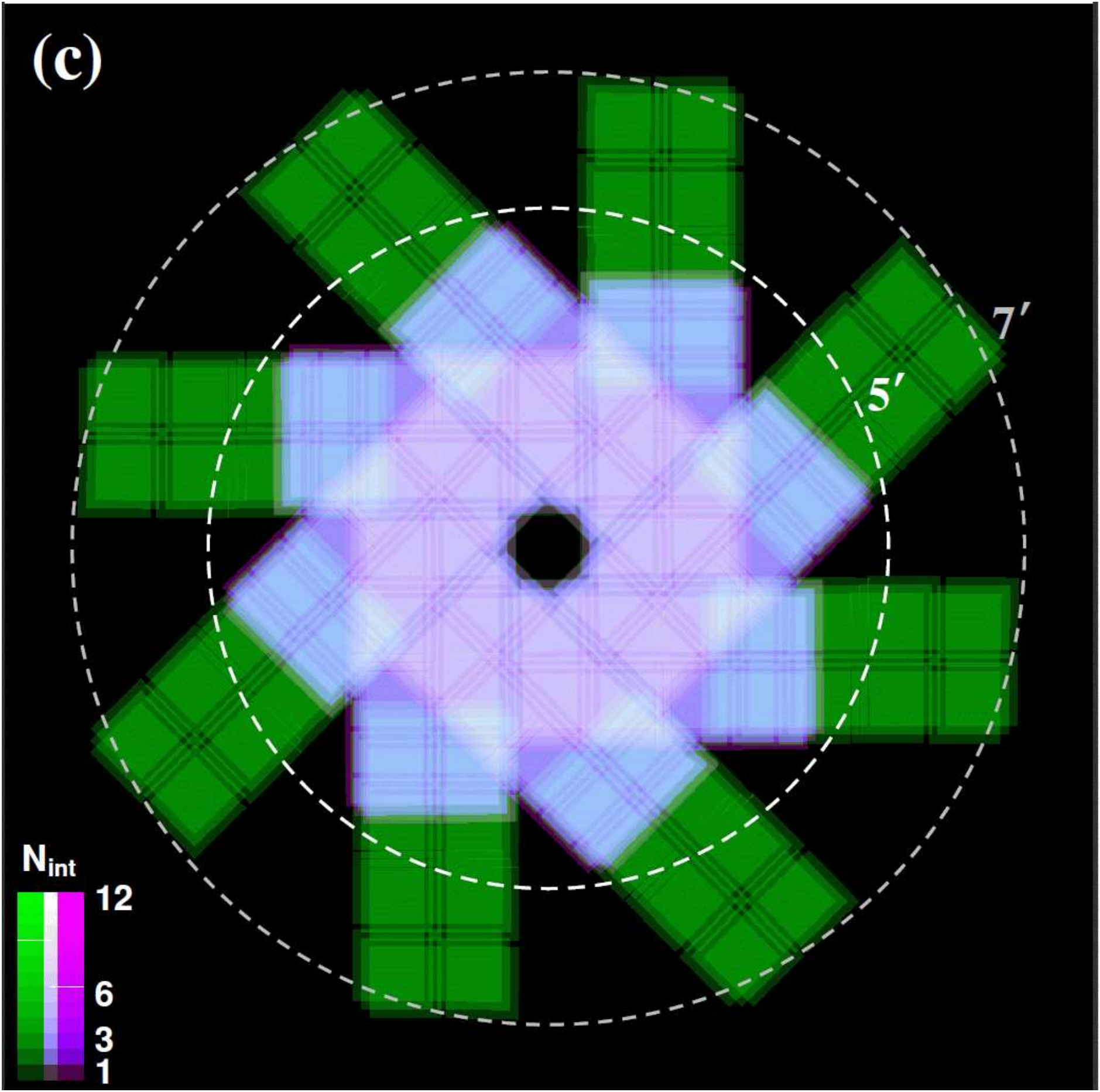}
  \includegraphics[height=0.245\txw]{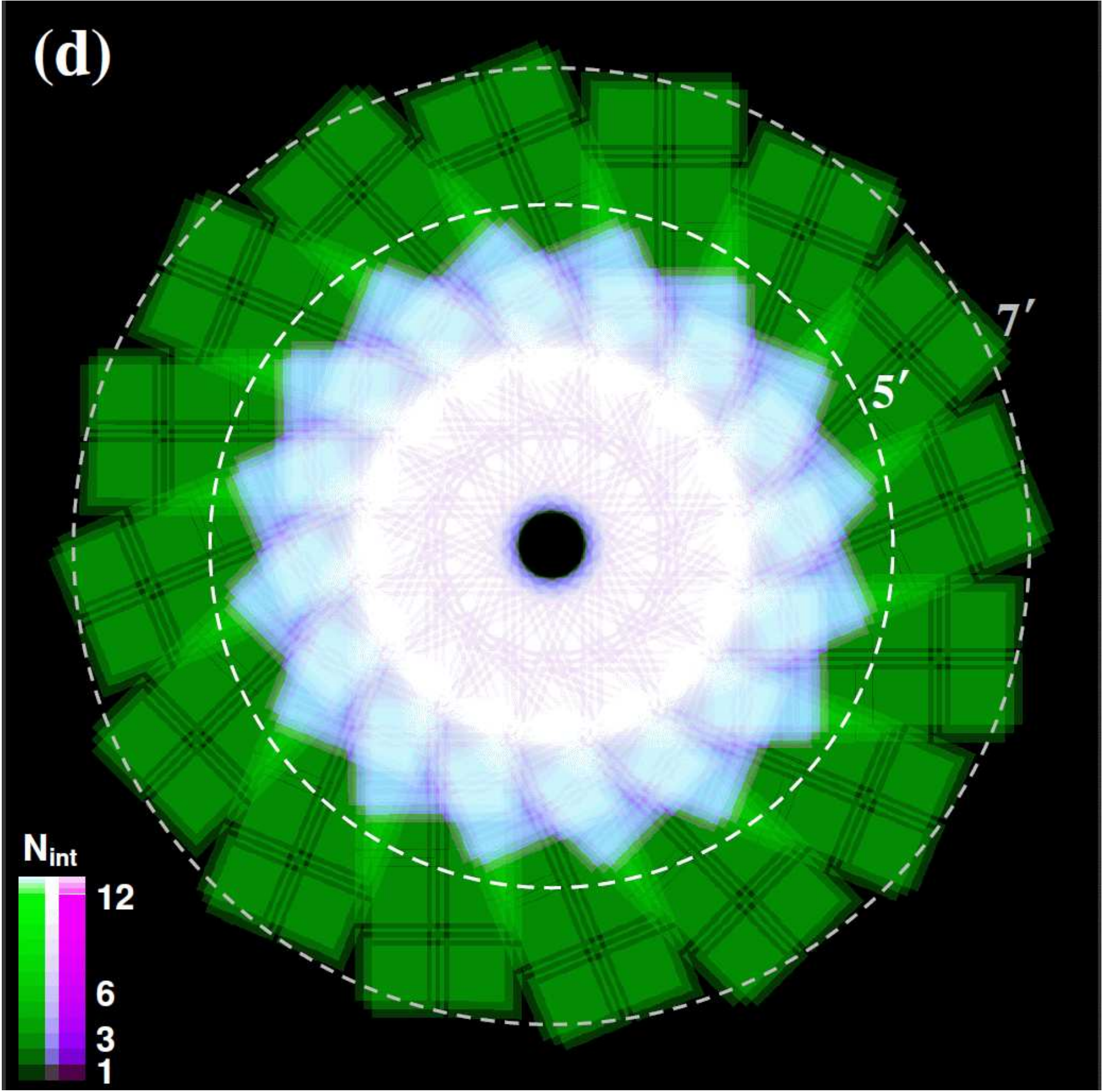}
}
\centerline{
  \rule{0.245\txw}{0pt}%-- empty dummy panel --%
  \includegraphics[height=0.245\txw]{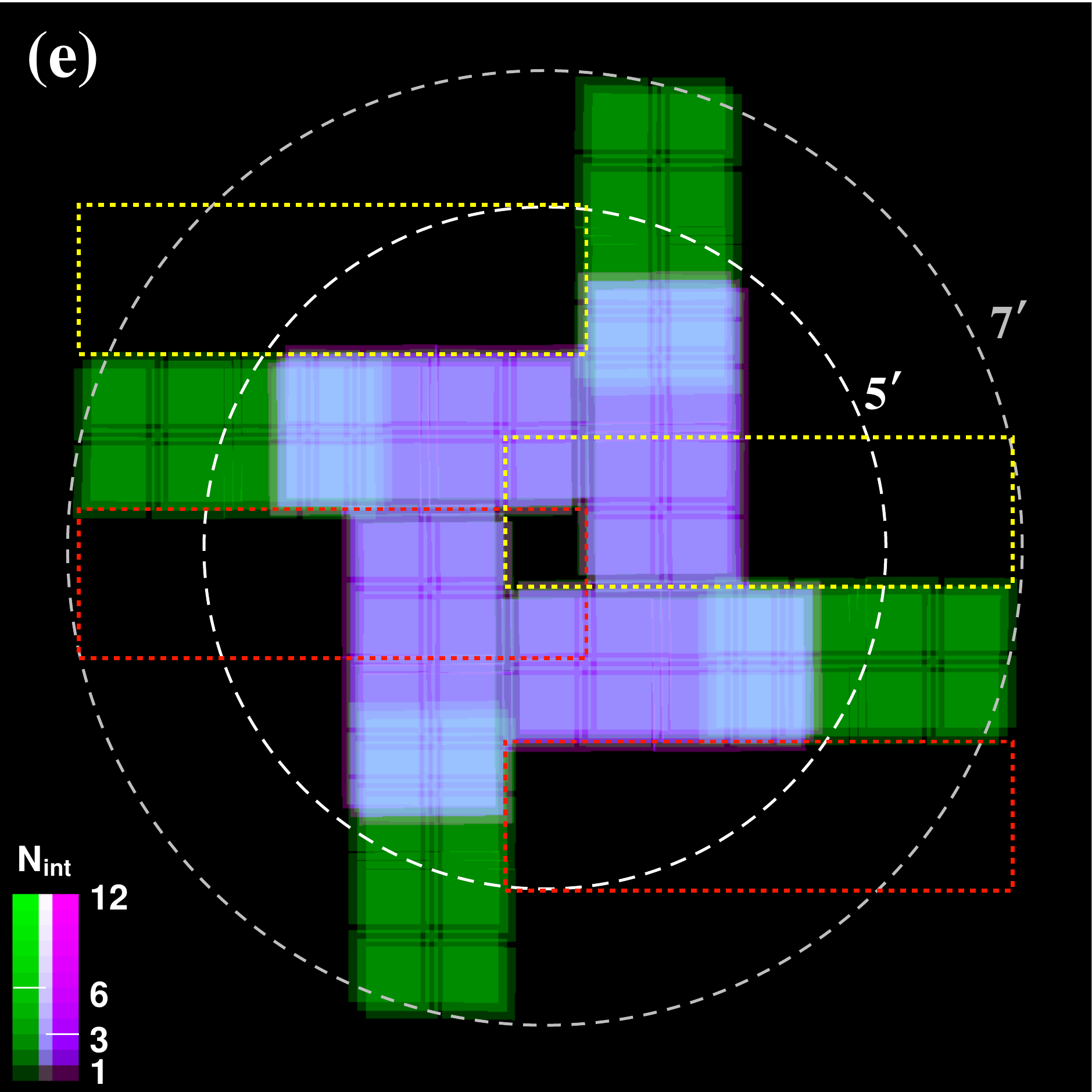}
  \includegraphics[height=0.245\txw]{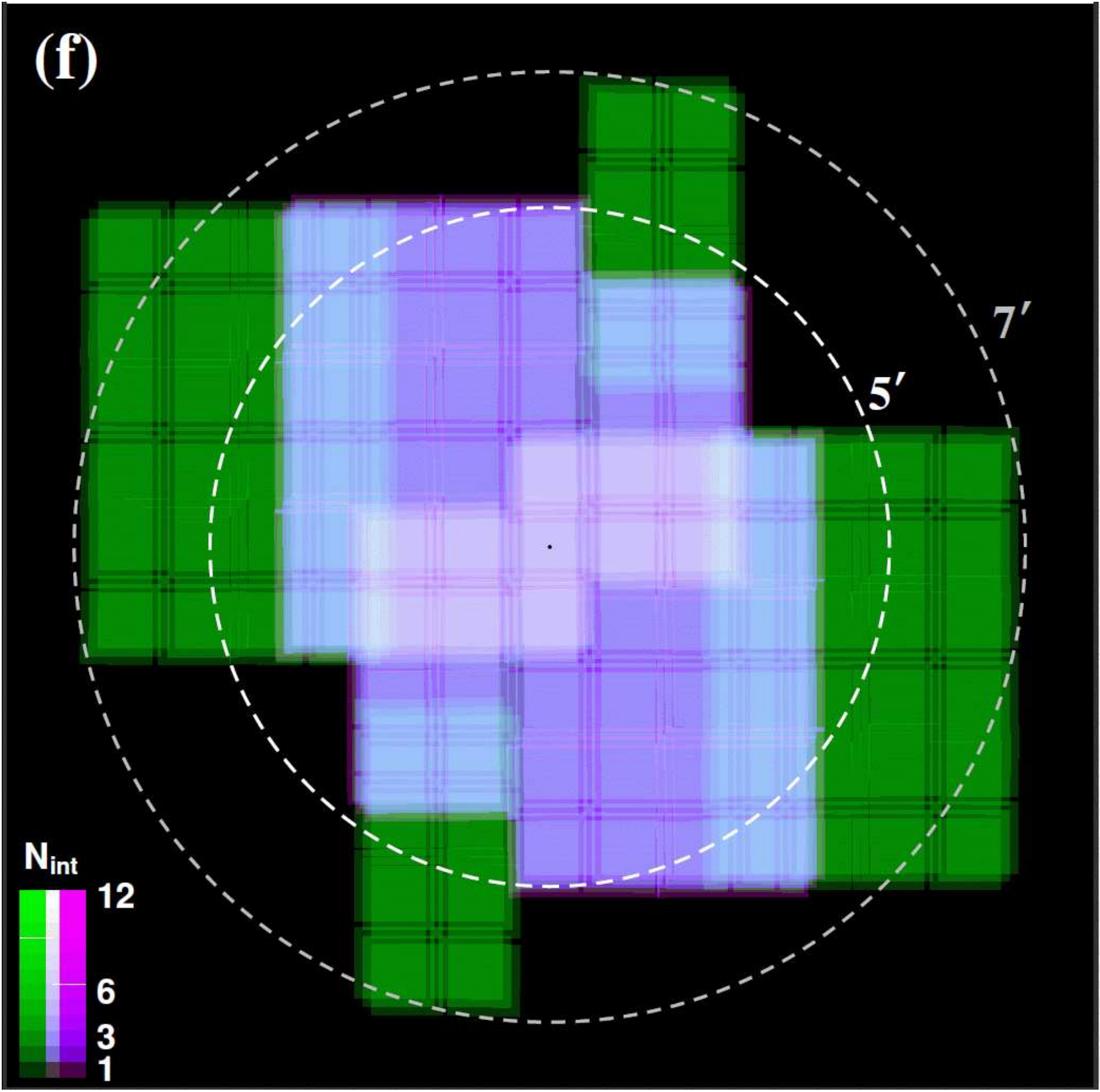}
  \includegraphics[height=0.245\txw]{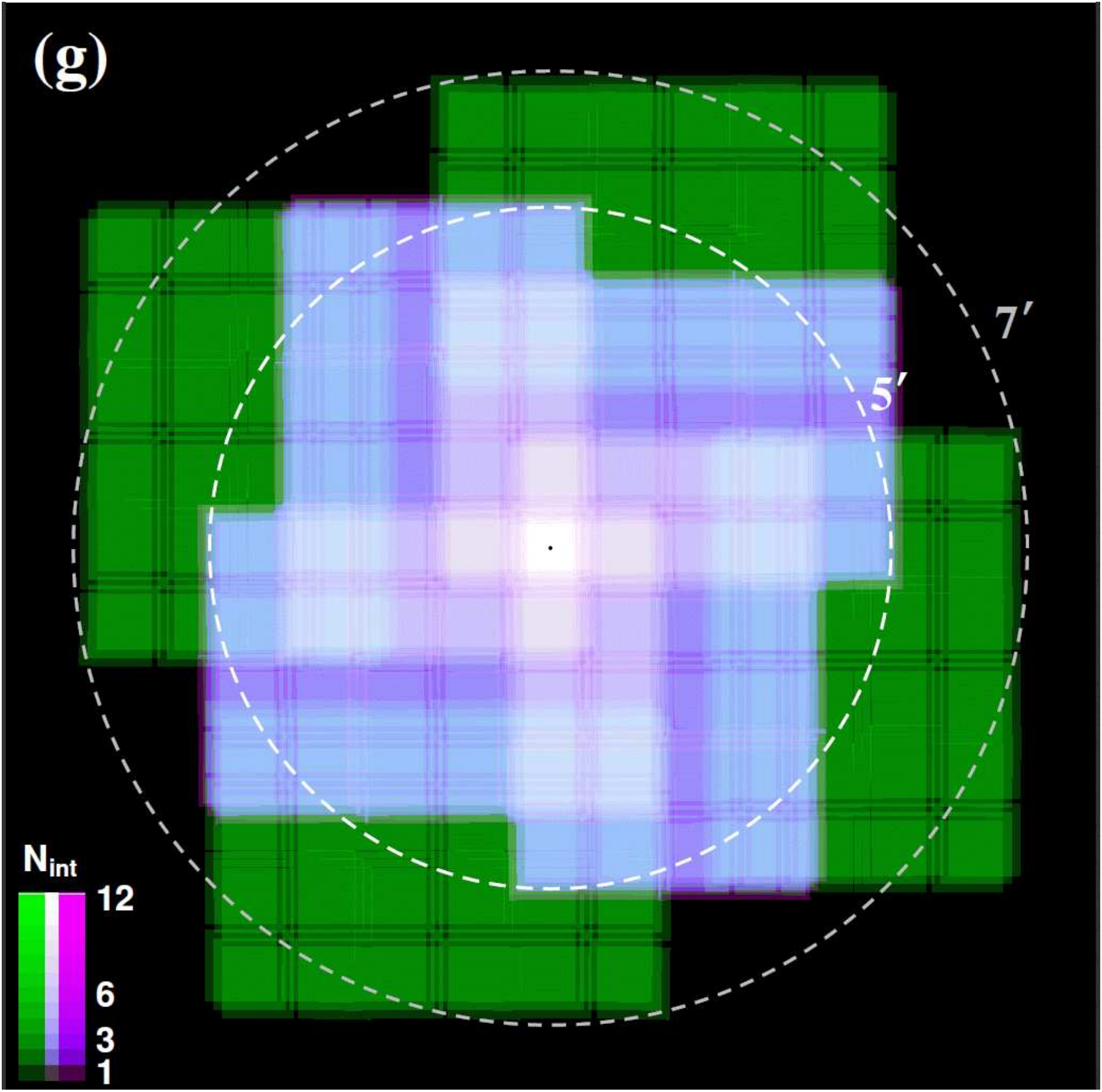}
}
\caption{\noindent\small
15$'$\ttimes15$'$ exposure maps of [\emph{a}] the \emph{unit visit} described
in \S~4.1, resulting in two, offset, contiguous areas covered by NIRCam and
NIRISS.  The nominal \JWST\ pointing center is indicated by a black dot.
[\emph{b}] a 4-spoke design like that adopted in GTO program 1176 of the
basic pattern of panel (\emph{a}), where the spokes have orientations that
differ by 90\tdeg\ on the sky.  Offsets are such that NIRCam covers the
NIRISS observations taken 180\,days ($\Delta$PA\,=\,180\tdeg) earlier or
later, and NIRISS will sample a large portion of the area imaged by
NIRCam.\deleted{\tsim63.1\%}
[\emph{c} and \emph{d}] Additional orientations ($\Delta$PA\,=\,45\tdeg and
\tpm22.5\tdeg, respectively) of similar pairs of spokes will increase the
total area covered, as well as the area sampled at multiple epochs.
[\emph{e}--\emph{g}] For the purpose of multi-filter object characterization
(as opposed to time-domain monitoring of objects already characterized), a
more efficient design could build directly on the four orientations of
GTO\,1176, maximizing the contiguous surface area covered per unit calendar
time.  The contiguous area of panel (\emph{g}) is nearly equivalent to the
\tsim153.9\,arcmin$^2$ of a circular area with a radius of 7$'$.
The actual orientation of these patterns on the sky, shown here aligned with
the cardinal axes, will depend on the \JWST\ launch date.
\label{fig:jwstNEPtds_expmaps}}
\end{figure*}
%%%%%%%%%%%%%%%%%%%%%%%%%%%%%%%%%%%%%%%%%%%%%%%%%%%%%%%%%%%%%%%%%%%%%%%%%%%%%

%\vspace*{-8pt}
\section{Possible Implementations}

Inspired by the very successful combination of \HST\ direct imaging and
grism spectroscopy, and in the spirit of the 3DHST, GRAPES, PEARS, FIGS,
GLASS, and WISP \HST\ grism surveys, we aim to find a \JWST\ survey
strategy that allows securing \emph{both} primary NIRCam imaging and
parallel NIRISS grism observations in the \JWST\ NEP TDF as part of a
\emph{single} observing program, while maximizing both the overall survey
area and the area of overlap between imaging and grism spectra.  We
consider two possible implementations ---for the purpose of
multi-wavelength 0.8--5.0\,\micron\ object characterization with or without
a time-domain component---, with an eventual goal of contiguous coverage of
an area of \tsim153.9\,arcmin$^2$, the area of a circular field with a
diameter of {14\tmin}.

%\vspace*{-2pt}
\subsection{A `Unit Visit' of NIRCam and NIRISS Observations}

After some experimentation with notional layouts and with supported
observation descriptions in \code{APT}, one natural pattern for a
\emph{unit visit}\footnote{With \emph{unit visit} we mean a sequence of
pointings and dithers executed within a single \JWST\ visit for a specified
set of optical elements.} emerged: it results in two, offset, contiguous
areas covered by NIRCam and NIRISS in parallel, and maximizes the potential
overlap of the footprints of the two instruments.  That pattern, shown in
Fig.~\ref{fig:jwstNEPtds_expmaps}\emph{a}, consists of a 2-point mosaic
with a pointing offset approximately equal to the width of a NIRISS
detector projected onto the sky, and is specified in \code{APT} as a
2\ttimes1 (columns\,\ttimes\,rows) mosaic with an overlap of 57\% of the
two NIRCam footprints (\ie both modules combined).  In combination with a
standard $\ge$3-point \swkey{INTRAMODULE} NIRCam dither at each mosaic
pointing, this unit pattern will fill both the large intra-module gap and
the small gaps between the four individual detectors in the SW channels of
each NIRCam module.  By specifying an offset of (190\farcs0,
$-$105\farcs0), the unit pattern becomes rotationally symmetric around the
desired nominal field center (indicated by the black dot in
Fig.~\ref{fig:jwstNEPtds_expmaps}\emph{a}), with maximal overlap between
the NIRISS and NIRCam coverages after rotation over {180\tdeg}.

%%%%%%%%%%%%%%%%%%%%%%%%%%%%%%%  TABLE 1  %%%%%%%%%%%%%%%%%%%%%%%%%%%%%%%%%%
%\placetable{1}
\noindent\begin{table}[b]
\caption{\small\rule{0pt}{0.4cm}Areal coverage vs.\ relative depth for the
	adopted Unit Visit, a 2\ttimes1 mosaic with a 3-point NIRCam
	\swkey{INTRAMODULE} dither\label{tab:table1}}
\setlength{\tabcolsep}{9pt}
\begin{tabular}{lrlcrl}
\hline\\[-16pt]
Instrument &     Depth     &          Area         & &
	      Depth    &          Area        \\[-4pt]
           &[$N_{\rm int}$]&$\!\!\!\!$[arcmin$^2$] & &
	[$N_{\rm int}$]&$\!\!\!\!$[arcmin$^2$]\\
\hline\\[-14pt]
NIRCam     &      1$^a$    &   2.240    & &    $>$0~~~   &   18.335  \\[-5pt]
           &      2~~~     &   3.951    & &    $>$1~~~   &   16.095  \\[-5pt]
           &      3$^b$    &   8.524    & &    $>$2$^c$  &   12.145  \\[-5pt]
           &      4~~~     &   0.926    & &    $>$3~~~   &  ~~3.621  \\[-5pt]
           &      5~~~     &   1.416    & &    $>$4~~~   &  ~~2.695  \\[-5pt]
           &      6~~~     &   1.279    & &    $>$5~~~   &  ~~1.279  \\[-2pt]
NIRISS     &      1$^a$    &   1.666    & &    $>$0~~~   &   11.580  \\[-5pt]
           &      2~~~     &   1.572    & &    $>$1~~~   &  ~~9.914  \\[-5pt]
           &      3$^b$    &   8.212    & &    $>$2$^c$  &  ~~8.342  \\[-5pt]
           &      4~~~     &   0.130    & &    $>$3~~~   &  ~~0.130  \\
\hline\\[-8pt]
\end{tabular}
\begin{minipage}{0.475\txw}{\small
Notes: (\emph{a}) at single-exposure depth, some unrecoverable image
defects will persist, and the PSF will be poorly sampled;
(\emph{b}) nominal, 3-dither depth is reached; (\emph{c}) nominal depth
is reached or exceeded.}
\end{minipage}
\end{table}
%%%%%%%%%%%%%%%%%%%%%%%%%%%%%%%%%%%%%%%%%%%%%%%%%%%%%%%%%%%%%%%%%%%%%%%%%%%%
  %
This unit pattern also results in an area of overlap between the coverage
of NIRCam Modules A and B (\tsim3.6\,arcmin$^2$), with an effective
exposure time up to twice the nominal one (see Table.~\ref{tab:table1} and
the dark shaded region in Fig.~\ref{fig:jwstNEPtds_expmaps}\emph{a}).  This
overlap allows photometric and astrometric cross-calibration between the
two modules, verification of systematics near the detection limit in areas
with nominal depth, as well as time-domain sampling on time-scales of
\tsim0.4--1.0\,hr (assuming a medium-deep NIRCam imaging survey to
\mAB\,\tsim\,29\,mag in 8 filters with coordinated parallel NIRISS grism
observations, and all the operational constraints imposed by the
observatory).  
An additional benefit as originally envisioned was that this pattern would
provide repeat sampling on time-scales of several hours for the area of
overlap between the two NIRCam modules.  This required cycling through each
of the filters at a given pointing before executing the offset in the
mosaic.  Overriding concerns for the longevity of the (large) NIRCam filter
wheels, however, disallow such cycling and force all pointing offsets in a
visit to be executed before moving to the next filter.

\vspace*{8pt}
\subsection{An IDS GTO Program to Start Object Characterization and
	Time-Domain Monitoring of the {\itshape JWST} NEP TDF}

Building on this unit visit pattern, program GTO-1176 (PI: R.~Windhorst)
will map an area of \tsim71.8\,arcmin$^2$ in the \JWST\ NEP TDF in 8
filters that span the 0.8--5.0\,\micron\ wavelength range of NIRCam. The
nominal 3-dither depth of \mAB\,\tsim\,29\,mag (5\,$\sigma$) is met or
exceeded for a total area of \tsim49.3\,arcmin$^2$.  The GTO coverage,
shown in Fig.~\ref{fig:jwstNEPtds_expmaps}\emph{b}, consists of four
distinct ``spokes'', each oriented 90\tdeg\ apart, and is charged
\tsim47\,hrs (4\ttimes11.7\,hrs) of calendar time.  Since the \JWST\ CVZs
are the \emph{only} place in the sky where \JWST\ can revisit a target with
an aperture orientation rotated over 180\tdeg\ with respect to a prior
visit ---\,without a significant penalty in the form of an increased
Zodiacal foreground brightness\,---, this is the \emph{only} place in the
sky where coordinated parallel observations with a second instrument (\eg
NIRISS, MIRI) can be made to almost fully overlap such prior NIRCam
observations, and vice versa. For the Windhorst IDS GTO program, the
combination of 1.8--2.2\,\micron\ NIRISS wide-field slitless spectroscopy
and NIRCam imaging for the purpose of source characterization was deemed
particularly powerful.  The areas where NIRISS and NIRCam coverage overlap
appear in light blue to purplish hues within the white dashed inner circle
of radius 5\tmin\ in Fig.~\ref{fig:jwstNEPtds_expmaps}\emph{b}.  The direct
images bracketing the dispersed NIRISS grism exposures reach to $>$29\,mag
as well, and allow time-domain monitoring of an area of
\tsim45.0\,arcmin$^2$ (of which \tsim33.6\,arcmin$^2$ at nominal depth or
better) on time-scales of 180 days when compared to the corresponding
NIRCam 2.0\,\micron\ images.

GO extensions of the GTO pattern with an additional four spokes with the
pattern rotated over 45\tdeg, as shown in
Fig.~\ref{fig:jwstNEPtds_expmaps}\emph{c}, would cover
\tsim114.4\,arcmin$^2$, while a further eight spokes at orientations of
$\Delta$PA\,=\,\tpm22.5\tdeg\ (Fig.~\ref{fig:jwstNEPtds_expmaps}\emph{d})
would approach our 153.9\,arcmin$^2$ goal. Such extensions with similar
pairs of spokes will not only increase the total areal coverage, but also
the area sampled at multiple epochs with increasingly dense time-domain
sampling. Furthermore, assuming NIRISS is used in parallel to NIRCam for
grism spectroscopy also in such GO programs, the number of distinct
orientations aids in disentangling spectra contaminated by signal from
neighboring objects within the field of view \citep[\eg][]{Ryanetal18}.

Once source characterization is available, the same unit visit pattern can
be used for efficient time-domain monitoring (to the same flux limits) of
sources that either move or vary in brightness.  For monitoring, one would
need only one SW (\eg F200W) and one LW (\eg F356W) NIRCam filter, observed
simultaneously, and direct imaging (\eg F200W) with NIRISS in parallel.

\subsection{Efficient Areal Mapping for Object Characterization}

When object characterization and time-domain monitoring are considered as
two entirely separate goals, an alternative, more efficient design using
the same unit visit pattern is possible for object characterization. It can
cover \tsim150--155 arcmin$^2$ in the \JWST\ NEP TDF in 8 NIRCam filters to
the same depth of \mAB\,\tsim\,29\,mag, with the same parallel NIRISS
1.8--2.2\,\micron\ grism spectroscopy and F200W direct imaging.  This
design is illustrated in
Fig.~\ref{fig:jwstNEPtds_expmaps}\emph{e}--\emph{g}.  Whereas the survey
strategy of \S~4.2 (Fig.~\ref{fig:jwstNEPtds_expmaps}\emph{b}--\emph{d})
rotates the pattern to leverage the year-round accessibility of the field
for time-domain science, it leaves a hole in the coverage at the nominal
field center.  An approach using offsets with respect to the spokes of
GTO-1176, and more constrained scheduling ---either within a few days from
the GTO visits, or a full year later--- could cover a contiguous area that
is nearly equivalent to the \tsim153.9\,arcmin$^2$ of a circular area with
a diameter of 14\tmin, while leaving no such hole.  In fact, the central
\tsim0.5, \tsim4.2, and \tsim26.3 arcmin$^2$ would be sampled at 4, 3, and
2 times the nominal depth, respectively. The coverage shown in
Fig.~\ref{fig:jwstNEPtds_expmaps}\emph{g} would require (\code{APT}25.4.4)
under 93\,hrs of calendar time in addition to the GTO observations (as
compared to an additional \tsim141\,hrs for the coverage of
Fig.~\ref{fig:jwstNEPtds_expmaps}\emph{d}).

\vspace*{8pt}

While the main aim of this design is to secure source characterization
in the NEP TDF early-on in the \JWST\ mission, even here there are
significant areas of overlap between the offset spokes that would provide
time-domain sampling on time-scales of \lsim3 days and of multiples of
\tsim90 days (up to \tsim1\,yr).

\section{Summary}

We described the selection of a new extragalactic survey field optimized
for time-domain science with \JWST. It is located within \JWST's northern
continuous viewing zone and is centered at (RA,\,Dec)$_{\rm J2000}$ =
(17:22:47.896,\,+65:49:21.54). This \JWST\ North Ecliptic Pole (NEP)
Time-Domain Field (TDF) is the \emph{only} \gsim14$'$ diameter area where
\JWST\ can observe a clean extragalactic survey field at any time of the
year and at arbitrary orientation, while leveraging \JWST's capability to
perform parallel science observations.  The NEP TDF will be targeted by
\JWST\ GTO program 1176, has a rich and growing complement of
multi-wavelength ancillary ground- and space-based observations, and has an
unmatched potential as a \JWST\ time-domain \emph{community field}.  We
estimated the number of sources expected to be detectable to
\mAB\,\tsim\,29\,mag with \JWST\ in the near-IR and to \muJy\ flux levels
in deep VLA 3--5\,GHz radio observations, as well as the subsets thereof
expected to show significant variability.  Last, we presented an efficient
unit visit, comprising primary NIRCam imaging and parallel NIRISS slitless
grism spectroscopy, with which observing programs can be designed for
wide-area source characterization and time-domain monitoring.  We encourage
\JWST\ GO proposers to adopt this efficient mode of observations for future
\JWST\ cycles, and hope that the NEP TDF will become one of \JWST's
community fields.

\acknowledgments

{Acknowledgements.
The authors acknowledge support from \JWST\ grants NAGS-12460, NNX14AN10G,
and 80NSSC18K0200 from the National Aeronautics and Space Administration
(NASA) Goddard Space Flight Center (GSFC), and from grant
HST-GO-15278.001-A from the Space Telescope Science Institute, which is
operated by the Association of Universities for Research in Astronomy, Inc.
(AURA) under NASA contract NAS\,5-26555.
We are grateful to the \HST\ Project at NASA GSFC and STScI for implementing
\JWST\ parallels also for science observations.

We thank Teresa Ashcraft for the preliminary reduction of the LBT/LBC data
used for Fig.~\ref{fig:jwstNEPtds_lbc}; Christopher Willmer for sharing his
Fortran code to compute the exposure map shown in
Fig.~\ref{fig:jwstNEPtds_expmaps}\emph{a}; Russell Ryan for kindly providing
number counts of Galactic stars computed specifically for the \JWST\ NEP TDF;
and Seth Cohen for suggesting a pinwheel pattern for \JWST\ NIRCam+NIRISS
observations.
We thank Walter Brisken, James Condon, Bill Cotton, Ken Kellermann,
Rick Perley, Peter Maksym, Norman Grogin, Patricia Royle, Anton Koekemoer,
Steven Rodney, Adam Riess, Lou Strolger, Nimish Hathi, Michael Rutkowski,
Chris Conselice, Simon Driver, Steven Finkelstein, Brenda Frye, Haojing Yan,
Dan Coe, Madeline Marshall, Victoria Jones, Cameron White,
G{\"u}nther Hasinger, Christopher Waters, Esther Hu, Giovanni Fazio,
Matt Ashby, Heidi Hammel, Stefanie Milam, Chad Trujillo, Hugo Messias,
Sangeeta Malhotra, Ian Smail, Myungshin Im, Ken Duncan, Justin Roberts-Pierel,
Andrea Ferrara, Stephen Wilkins, Michelle Edwards, Volker Tolls,
Mehmet Alpaslan, Linhua Jiang, Patrick Kelly, Nor Pirzkal, Martin Ward,
Jennifer Patience, Michael Line, Silvia Bonoli, Renato Dupke, Vik Dhillon,
and Michele Bannister for useful discussions and (proposals for) ancillary
observations in the \JWST\ NEP TDF, and/or for comments on the manuscript,
and Neill Reid, Ken Sembach, John Mather, and Eric Smith for inspiring us to
develop this field as a \JWST\ Community Field.  We thank the anonymous
referee for helpful comments and suggestions that improved this paper.

This publication used data products from \WISE, which is a joint project of
the University of California, Los Angeles, and the Jet Propulsion Laboratory
(JPL)/California Institute of Technology (Caltech), funded by NASA.  We used
Atlas Images from the Two Micron All Sky Survey (2MASS), a joint project of
the University of Massachusetts and the Infrared Processing and Analysis
Center (IPAC)/Caltech, funded by NASA and the National Science Foundation
(NSF).  We also used data products from the Sloan Digital Sky Survey (SDSS),
funded by the Alfred~P.~Sloan Foundation, the Participating Institutions,
NASA, NSF, the U.S.~Department of Energy, the Japanese Monbukagakusho, and
the Max Planck Society.  The SDSS is managed by the Astrophysical Research
Consortium (ARC) for the Participating Institutions: The University of
Chicago, Fermilab, the Institute for Advanced Study, the Japan Participation
Group, The Johns Hopkins University, Los Alamos National Laboratory, the
Max-Planck-Institutes for Astronomy (MPIA) and Astrophysics (MPA), New
Mexico State University, University of Pittsburgh, Princeton University, the
U.S.~Naval Observatory, and the University of Washington.
This research has made use of the NASA/IPAC Infrared Science Archive (IRSA)
and NASA/IPAC Extragalactic Database (NED), which are operated by JPL/Caltech
under contract with NASA.

Based in part\deleted{also} on observations (PI: R.A.~Jansen) made with
the Large Binocular Telescope (LBT).  The LBT is an international
collaboration among institutions in the United States, Italy, and Germany.
LBT Corporation partners are: The University of Arizona on behalf of the
Arizona university system; Istituto Nazionale di Astrofisica, Italy; LBT
Beteiligungsgesellschaft, Germany, representing the Max-Planck Society, the
Astrophysical Institute Potsdam, and Heidelberg University; The Ohio State
University; and The Research Corporation, on behalf of the Universities of
Notre Dame, of Minnesota and of Virginia.

We used \code{Montage} \citep{Montage05,Montage09}, funded by NSF
\deleted{under Grant Number ACI-1440620,}and by NASA's Earth Science
Technology Office, Computation Technologies Project, under Cooperative
Agreement\deleted{Number NCC5-626} between NASA and Caltech, and maintained
by the NASA/IPAC Infrared Science Archive.  \code{Montage} includes an
adaptation of the \code{MOPEX} algorithm developed at the Spitzer Science
Center, and software developed by IPAC for the US National Virtual
Observatory, which is sponsored by NSF.
We used \code{IRAF} \citep{IRAF1,IRAF2}, a general purpose software system
for the reduction and analysis of astronomical data, which is written and
distributed by the National Optical Astronomy Observatories, which are
operated by AURA\deleted{the Association of Universities for Research in
Astronomy, Inc.,} under cooperative agreement with NSF.
Images and color composites were rendered using SAOImage \code{DS9}
\citep{DS9}, developed by the Smithsonian Astrophysical Observatory, with
funding from the Chandra X-ray Science Center\deleted{(NAS8-03060)}, the
High Energy Astrophysics Science Archive Center\deleted{(NCC5-568)}, and
the JWST Mission office at Space Telesccope Science
Institute\deleted{(NAS-03127)}.\\[2pt]
This research has made use of NASA's Astrophysics Data System (ADS)
bibliographic services \citep{ADS}.\\

}

\software{\code{(C)FITSIO}, \code{DS9}, \code{IDL}, \code{IRAF},
	\code{Montage}, \code{SuperMongo}, \code{SCAMP}, \code{SWARP};
	Fortran77 code and utilities written by the lead author
	(\code{narg/arg}, \code{myfixnan}, \code{select}, \code{mkmask},
	\code{listpix})}

\facilities{Large Binocular Telescope (Large Binocular Cameras);
	Two Micron All Sky Survey; Wide-field Infrared Survey Explorer}

\end{document}